\documentclass[pra,letterpaper,aps,10pt,superscriptaddress,twocolumn,floatfix,showpacs]{revtex4-1}
\usepackage{graphicx}
\usepackage{amsmath}
\usepackage{amsfonts}
\usepackage{amssymb}
\usepackage{footmisc}
\usepackage{epsfig}
\usepackage[pdftex]{color}
\usepackage{amsmath,graphicx,amssymb,braket,xcolor,subfigure,upgreek}
\usepackage[colorlinks, linkcolor=blue, citecolor=blue, urlcolor=blue, breaklinks=true]{hyperref}
\usepackage{microtype}
\usepackage{bbm}
\usepackage{color}
\usepackage{braket}
\usepackage{amsmath,amssymb,amsfonts,latexsym,color,dcolumn,bm,amsbsy}
\usepackage[english]{babel}
\usepackage{graphicx}
\usepackage[utf8]{inputenc}
\usepackage{textcomp}
\usepackage{braket}

\newcommand{\D}{{\cal{D}}}
\newcommand{\Dd}{{\cal{D}^{\dagger}}}
\newcommand{\bd}{b^{\dagger}}
\newcommand{\ad}{a^{\dagger}}
\newcommand{\sd}{\sigma^{\dagger}}
\newcommand{\s}{\sigma}

\newcommand{\nuk}{\nu_{\text{k}}}
\newcommand{\bk}{b_{\text{k}}}
\newcommand{\bdk}{b^\dagger_{\text{k}}}

\newcommand{\binp}{b_{\text{in}}}
\newcommand{\binpd}{b_{\text{in}}^{\dagger}}
\newcommand{\siginp}{\sigma_{\text{in}}}
\newcommand{\siginpd}{\sigma_{\text{in}}^{\dagger}}

\newcommand{\T}{\textbf{T}}
\newcommand{\ome}{\omega_{\text{e}}}
\newcommand{\omc}{\omega_{\text{c}}}
\newcommand{\bomc}{\bar{\omega}_{\text{c}}}
\newcommand{\oml}{\omega_{\ell}}
\begin{document}

\title{Langevin Approach to Quantum Optics with Molecules}
\author{Michael Reitz, Christian Sommer and Claudiu Genes}
\affiliation{Max Planck Institute for the Science of Light, Staudtstra{\ss}e 2,
D-91058 Erlangen, Germany}
\date{\today}

\begin{abstract}
We investigate the interaction between light and molecular systems modeled as quantum emitters coupled to a multitude of vibrational modes via a Holstein-type interaction. We follow a quantum Langevin equations approach that allows for analytical derivations of absorption and fluorescence profiles of molecules driven by classical fields or coupled to quantized optical modes. We retrieve analytical expressions for the modification of the radiative emission branching ratio in the Purcell regime and for the asymmetric cavity transmission associated with dissipative cross-talk between upper and lower polaritons in the strong coupling regime. We also characterize the F\"{o}rster resonance energy transfer process between donor-acceptor molecules mediated by the vacuum or by a cavity mode.
\end{abstract}


\maketitle

Recent experimental progress~\cite{wang2019turning} has shown that the Purcell enhancement of the zero-phonon line of a single molecule can strongly alter the branching ratio of spontaneous emission between the line of interest and additional Stokes lines thus turning the molecule into an ideal quantum emitter. At the mesoscopic level, experiments in the collective strong coupling regime of organic molecules with cavities or delocalized plasmonic modes have shown important light-induced modifications of material properties. Experimental and theoretical endeavors go into the direction of charge and energy transport enhancement~\cite{orgiu2015conductivity,schachenmayer2015cavity,feist2015extraordinary,hagenmuller2017cavity,hagenmuller2018cavity}, F\"{o}rster resonance energy transfer (FRET) enhancement~\cite{zhong2016non,zhong2017energy,feist2017long,du2018theory,reitz2018energy}, modified chemical reactivity~\cite{hutchinson2012modifying,galego2016suppressing,herrera2016cavity,martinezmartinez2018can,kampschulte2018cavity}, polariton dynamics~\cite{schwartz2013polariton,durand2015non} etc. Oftentimes however, experiments rely on theoretical models developed for standard cavity quantum electrodynamics (cavity QED)~\cite{haroche1989cavity, berman1994cavity, walther2006cavity} with two-level systems where one distinguishes between i) the Purcell regime, characterized by modifications of the spontaneous emission rates and ii) the strong coupling regime leading to the occurrence of hybrid light-matter states referred to as polaritons. Recent theoretical efforts aim at covering this gap by solving a generalized light-electronic-vibrations problem modeled as a Holstein-Tavis-Cummings Hamiltonian. Investigations aim at providing an understanding of the vibrationally induced cavity polariton asymmetry~\cite{durand2015non,neuman2018origin}, vibrationally dressed polaritons~\cite{zeb2018exact}, dark vibronic polaritons ~\cite{herrera2017dark,herrera2018theory}, developing a cavity Born-Oppenheimer theory~\cite{flick2017atoms,flick2017cavityBorn} or deriving relevant simplified models for large scale numerics in the mesoscopic limit~\cite{pino2018tensor}.

\begin{figure*}[t]
\includegraphics[width=2.10\columnwidth]{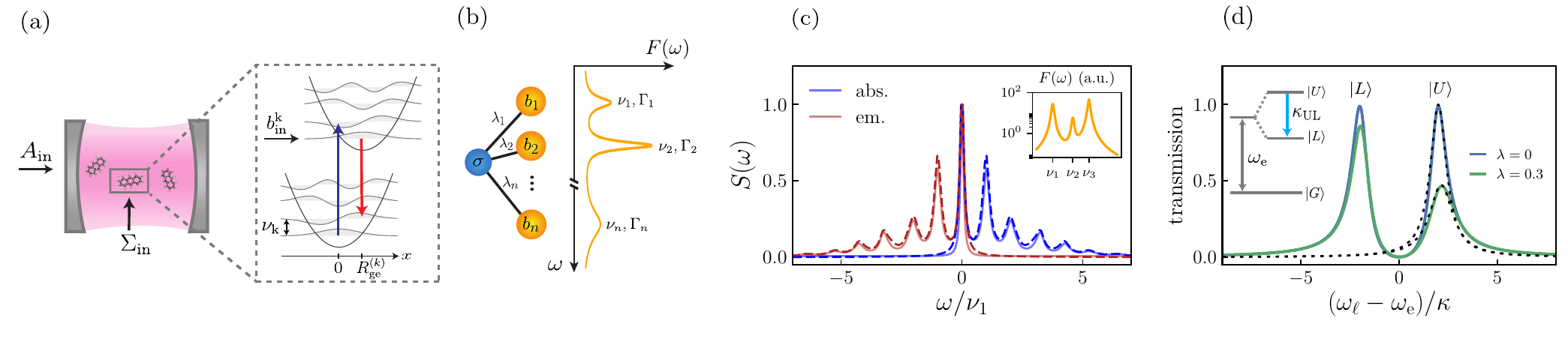}
\caption{(a) The equilibrium mismatch $R_{\text{ge}}^{(k)}$ between ground and excited electronic potential landscapes along a nuclear coordinate leads to the standard Franck-Condon physics with distinct optimal absorption (blue arrow) and emission (red arrow) lines. (b) \textit{Molecular box} consisting of an electronic transition operator $\s$ coupled to $n$ vibrations $b_{\text{k}}$ with frequencies, linewidths and strengths dictated by the spectral density function $F(\omega)$. (c) Absorption/emission profiles for a toy-model molecule with 3 vibrational modes (spectral density shown in inset). Solid lines show numerical master equation results, dashed lines are analytical results. (d) Cavity transmission $|\text{t}_\text{c}|^2$ on resonance $\omc=\ome$ showing polariton asymmetry for $\lambda=0.3$. Full lines are numerical simulations with $\Gamma=1.5\kappa$, $g=2\kappa$, $\nu=4\kappa$ while dotted lines are fitted Lorentzians with the analytically derived $\kappa_{\text{UL}}$.}
\label{fig1}
\end{figure*}

We provide here an alternative path based on solving the Holstein-Tavis-Cummings dynamics at the level of operators rather than states. The basic model considers a \textit{molecular box} (see Fig.~\ref{fig1}(b)) comprised of an internal electronic transition coupled to any number of vibrational modes. Radiative decay and vibrational relaxation are included as stochastic source terms in a set of coupled standard quantum Langevin equations~\cite{gardiner2004quantum, gardiner1985input, collett1984squeezing, breuer2002theory, walls2006quantumoptics} for vibrational and polaron operators (similarly applied in optomechanical systems~\cite{genes2008ground,rabl2011photon,nunnenkamp2011single}). The method is numerically tested showing correct predictions for molecular emission/absorption lines. In the strong coupling regime of cavity QED, polaron-polariton dynamics provides insight in the upper-lower polariton dissipative coupling. In the Purcell regime we analytically derive the modified branching ratio of radiative transitions between the zero-phonon line and all other Stokes lines~\cite{wang2019turning}. For different molecular species, we derive the donor--acceptor FRET rate occurring either via direct dipole-dipole virtual photon exchanges or in the presence of a cavity. We find regimes where the direct transfer can be increased by a Purcell-enhancement of the acceptor's absorption lines.\\

\noindent \textbf{Model} --- We consider $N$ molecules ($j$ running from $1$ to $N$) with ground $\ket{g_j}$ and excited $\ket{e_j}$ electronic states and Pauli lowering operators $\s_j=\ket{g_j}\bra{e_j}$. Each molecule's excited state orbital is perturbed by its coupling to $n$ vibrational modes (frequencies $\nu_\text{k}$), corresponding to stretching along different nuclear coordinates. These are described by creation/annihilation operators $\bd_{j\text{k}}$ and $b_{j\text{k}}$ ($k$ running from $1$ to $n$). The vibronic couplings (see Supplemental Material (SM) \cite{SM}) are modeled via the standard Holstein Hamiltonian~\cite{holstein1959study} (with $\hbar=1$) with coupling parameters $\lambda_{\text{k}}=\mu_{\text{k}}\nu_{\text{k}} R^{(k)}_\text{ge}r^{(k)}_\text{zpm}$ (the square root of the Huang-Rhys factor) where the spread of the zero point motion wavepacket is $r^{(k)}_\text{zpm}=\sqrt{1/(2\mu_{\text{k}}\nu_{\text{k}})}$ (with $\mu_{\text{k}}$ -- the effective mass for the given $\text{k}$-vibration). The free dynamics of a single molecule is then governed by
\begin{equation}
h^{(j)}_0= \left( \ome + \sum_{\text{k}=1}^{n} \lambda_{\text{k}}^2 \nu_{\text{k}} \right) \sd_j \s_j  +   \sum_{\text{k}=1}^{n}\nu_\text{\text{k}} \bd_{j\text{k}} b_{j\text{k}}.
\label{h0j}
\end{equation}
Holstein contributions from all vibrational modes add as
\begin{equation}
h^{(j)}_H =  -   \sum_{\text{k}=1}^{n} \lambda_{\text{k}} \nu_{\text{k}} \left(\bd_{j\text{k}}+b_{j\text{k}}\right) \sd_j \s_j,
\label{hHj}
\end{equation}
and lead to the standard Franck-Condon physics illustrated in Fig.~\ref{fig1}(a). Notice that for each molecule one could define a single collective vibrational \textit{bright} mode that couples to the electronic degrees of freedom and diagonalize the free vibrational part in terms of \textit{dark} modes orthogonal to it and bright-dark couplings. In some cases such chain transformation (as used in Ref.~\cite{pino2018tensor}) can provide a great simplification to the problem especially for numerical simulations. This is not necessary in our treatment as we can include all vibrational modes on equal footing at the analytical level. The molecules can be embedded in a cavity (see Fig.~\ref{fig1}(a)) with a single mode of interest - annihilation operator $a$ with frequency $\omc$. Each molecule is coupled with $g_j$ to the cavity field; the interaction is the standard Tavis-Cummings Hamiltonian
\begin{equation}
H_c = \omc \ad a + i \sum_{j=1}^N \left( g_j \ad \s_j-g^{*}_j  \sd_j a \right).
\label{Hc}
\end{equation}
At high densities, molecules sit in the near-field of their neighbors which leads to vacuum mediated dipole-dipole exchanges modeled as $h^{(ij)}_{dd}=\Omega_{ij}\s_i \sd_{j}$ (with a strongly distance dependent exchange rate $\Omega_{ij}$). In addition, either the molecules or the cavity can be driven by a laser at frequency $\omega_{\ell}$ with pump amplitude $\eta_{\ell}$ (or $\eta_{\text{c}}$ - for cavity pumping). Adding all the parts together, the system coherent dynamics is governed by
\begin{equation}
H = \sum_{j=1}^N \left(h^{(j)}_0+h^{(j)}_H\right)+\sum_{i\neq j}^{N}h^{(ij)}_{dd}+ H_c+H_{\ell}.
\label{Hc}
\end{equation}
Laser driving is included as $H_{\ell}=i\eta_{\ell} \sum_{j=1}^{N}\left(\sd_j e^{-i\omega_{\ell}t}-\s_j e^{i\omega_{\ell}t}\right)$ (for molecule pumping with amplitude $\eta_{\ell}$) or as $H_{\ell}=i\eta_{\text{c}} \left(\ad e^{-i\omega_{\ell}t}-a e^{i\omega_{\ell}t}\right)$ (with $\eta_{\text{c}}$ for cavity pumping). Dissipation is added as Lindblad terms: radiative decay (at rates $\gamma_j$ with collapse operators $\s_j$), cavity field decay (rate $\kappa$, collapse operator $a$) and internal vibration relaxation (rates $\Gamma_{jk}$, collapse operators $b_{jk}$). For a collapse operator $\cal{O}$ with rate $\gamma_{\cal{O}}$ the Lindblad term applied to a density operator $\rho$ is: ${\cal{L}}_{\cal{O}}[\rho]=\gamma_{\cal{O}}\left\{2\cal{O}\rho \cal{O}^\dagger-\rho\cal{O}^\dagger\cal{O}-\cal{O}^\dagger\cal{O}\rho\right\}$. For the optical cavity field the Lindblad term is exact even in the zero temperature limit. For vibrations, this is an approximation as the physics of bulk-vibrations interactions could be better included as a Brownian noise dissipation model~\cite{hu1992quantum}. We compare the two dissipation models in the SM \cite{SM} showing agreement for spectral line frequencies and identifying terms leading to disagreement in linewidth predictions.\\

\noindent \textbf{Single molecule spectroscopy}--- Setting $g=0$ and assuming a single molecule $N=1$ coupled to many vibrational modes $n>1$, the molecular Hamiltonian can be diagonalized by a standard (level dependent) polaron transformation $U^{\dagger}=(\Dd)^{\sigma^\dagger \sigma}$. This is a collective transformation as $\Dd=\D^\dagger_{1}...\D^\dagger_{n}$ where each individual displacement is defined as $\D_\text{k}=\exp(-i\sqrt{2}\lambda_\text{k} p_\text{k})=\exp[\lambda_\text{k} (\bdk-\bk)]$ with the momentum quadrature  $p_\text{k}$ as a generator. The driving part is transformed accordingly $\tilde{H}_{\ell}=i\eta_\ell\left(\s \Dd {e^{i\omega_{\ell}t}}-\D \sd e^{-i\omega_{\ell}t}\right)$ and describes the generation of a superposition of coherent wavepackets with amplitudes $\lambda_\text{k}$ -- accompanying stimulated absorption and inverse displacement (with $-\lambda_\text{k}$) for stimulated emission. However, we only indirectly make use of this transformation by shifting to a Langevin description where the time dynamics of a \textit{polaron operator} $\tilde{\s}(t)=\s (t)\Dd(t)$ is followed. Under weak driving conditions $\eta_{\ell}\ll\gamma$ the exact Langevin equations (see SM \cite{SM}) can be simplified to:
\begin{subequations}\label{eq:Lang1}
\begin{align}
\frac{d b_\text{k}(t)}{dt} &= -(\Gamma_{\text{k}}+i\nu_\text{k})\bk(t) +\sqrt{2\Gamma_\text{k}}\binp^\text{k}(t),\\
\frac{d\tilde{\s}(t)}{dt} &= -\left[\gamma-i(\omega_{\ell}-\ome)\right]\tilde{\s}(t)+\sqrt{2\gamma} \Dd(t) \Sigma_\text{in}(t),
\end{align}
\end{subequations}
where we condensed the noise affecting the electronic transition into a term of non-vanishing average $\Sigma_\text{{in}}(t)=\eta_\ell/\sqrt{2\gamma}+\siginp(t)$. For zero temperature $T=0$, the only non-zero noise correlations are $\braket{\siginp(t)\s_{\text{in}}^\dagger(t')}=\delta(t-t')$. The reservoir model is consistent with models used elsewhere~\cite{neuman2018origin}: the power spectrum of the vibrational force $F=\lambda \nu (b+\bd)$ is given by a Lorentzian centered around the vibrational frequency $F(\omega)= 2\lambda^2 \nu^2 \Gamma/\left[\Gamma^2+(\omega-\nu)^2\right]$. For simplicity of presentation we will proceed with $\text{T}=0$; however the SM \cite{SM} presents a general approach to include thermal effects by including a finite $\text{T}$ via $\braket{b_{\text{in}}^\dagger(t)b_{\text{in}}(t')}=\Gamma \bar{n} \delta(t-t')$ and $\braket{b_{\text{in}}(t)b_{\text{in}}^\dagger(t')}=\Gamma (\bar{n}+1)\delta(t-t')$ where the thermal occupancy is given by $\bar{n}=[\exp({\hbar \nu/(k_\text{B}\text{T})})-1]^{-1}$.
The first equation can be integrated to give a solution for $\D(t)$ in terms of vibrational input noises $\binp^\text{k}(t)$ and $b_{\text{in}}^{\text{k},\dagger}(t)$. We can then derive the solution for $\tilde{\s}(t)=\s (t)\Dd(t)$ and subsequently for the bare Pauli operator
\begin{align}
\label{pauli}
\s(t) &=\s(0)\D(t)\Dd(0)e^{-\left[\gamma-i(\oml-\ome)\right]t} \\
&+ \sqrt{2\gamma}\int_{0}^t dt' e^{-\left[\gamma-i(\oml-\ome)\right](t-t')}\Sigma_\text{{in}}(t')\D(t)\Dd(t').\nonumber
\end{align}

\noindent This expression allows for the derivation of any system property both in the transient regime as well as in steady state. For the absorption (extinction) profile, we evaluate $\braket{\sd(t)\s(t)}$ in steady state under continuous driving conditions (see SM \cite{SM}). As the input noises commute at any time, one is left with the task of evaluating $\braket{\D_\text{k}(t)\D^\dagger_\text{k}(t')}=\exp(-\lambda^2_\text{k})\exp\left(\lambda_\text{k}^2[\exp(-\left(\Gamma_\text{k}+i\nu_\text{k})(t-t')\right)]\right)$. After the subsequent integration the excited state population $P^{\text{ss}}_\text{e}=\braket{\sd\s(t)}$ is given by
\begin{align}
P^{\text{ss}}_\text{e}=\eta_{\ell}^2\sum_{{{\left\{m_\text{k}\right\}}}={0}}^{\infty} \frac{\text{s}_{m_1}^{\lambda_1}...\text{s}_{m_{n}}^{\lambda_n}\left[\gamma + \Gamma_{\left\{m_\text{k}\right\}}\right]/\gamma}{\left[\gamma + \Gamma_{\left\{m_\text{k}\right\}}\right]^2 +\left[\oml-\ome - \nu_{\left\{m_\text{k}\right\}}\right]^2}.
\end{align}
The sum above runs over all the indices $\left\{m_\text{k}\right\}\equiv{m_1,...,m_n}$ and the numerator contains a product of Poissonian coefficients $\text{s}_{m_\text{k}}^{\lambda_\text{k}}=\exp({-\lambda_\text{k}^2})\lambda_\text{k}^{2m_\text{k}}/{m_\text{k}}!$. The denominator indicates a series of resonances corresponding to frequency up-shifted absorption lines with vibrationally modified linewidths $\Gamma_{\left\{m_\text{k}\right\}}=\sum_{\text{k}=1}^{n}m_\text{k} \Gamma_\text{k}$ and corresponding vibrationally-induced energy shifts $\nu_{\left\{m_\text{k}\right\}}=\sum_{\text{k}=1}^{n}m_\text{k}\nu_\text{k}$ (see Fig.~\ref{fig1}(c)). Notice that the absorption profile in the pure two-level system limit is immediately recovered by setting all $\lambda_\text{k}=0$, which implies that only terms with all $m_\text{k}=0$ contribute such that the expected simple Lorentzian is obtained $\braket{\sd\s(t)}=\eta_{\ell}^2 /[\gamma^2+(\oml-\ome)^2]$.\\
\noindent To evaluate the fluorescence spectrum (see SM \cite{SM}), we assume initial preparation of the molecule in a mixed state between ground and excited electronic states with no vibrations and compute the two-time correlations $\braket{\sd(\tau)\s(0)}$. The spectrum is defined as $S_\text{em}(\omega)=2\Re{\int_{0}^{\infty}d\tau \braket{\sd(\tau)\s(0)}e^{-i\omega\tau}}$ \cite{auffeves2008spontaneous} and is expressed as

\begin{align}
S_\text{em}(\omega)=\sum_{{{\left\{m_\text{k}\right\}}}={0}}^{\infty} \frac{2\text{s}_{m_1}^{\lambda_1}...\text{s}_{m_n}^{\lambda_n}\left[\gamma + \Gamma_{\left\{m_\text{k}\right\}}\right]}{\left[\gamma + \Gamma_{\left\{m_\text{k}\right\}}\right]^2 +\left[\omega-\ome +\nu_{\left\{m_\text{k}\right\}}\right]^2}.
\end{align}

\noindent The Stokes lines are present in the resonances of the denominator requiring $\omega=\ome-\nu_{\left\{m_\text{k}\right\}}$ (see Fig.~\ref{fig1}(c) - plotted in a frame rotating with $\oml$). Setting all $m_\text{k}=0$ reveals the zero-phonon line strength in the emission proportional to $\text{s}_{0}^{\lambda_1}...\text{s}_{0}^{\lambda_n}=e^{-\left(\lambda_1^2+...+\lambda_n^2\right)}$. Notice that for a molecule with many vibrations, even if all Huang-Rhys factors are small, the collective vibrationally induced reduction of the oscillator strength on the zero-phonon line can be considerably large. Also, the branching ratio of the molecules defined as $\alpha=\gamma_{\text{00}}/(\gamma_{\text{00}}+\gamma')$ ($\gamma_{\text{00}}$ is the radiative rate of the zero-phonon line and $\gamma'$ represents the sum of rates to all other states in the ground state manifold) can be immediately estimated to  $\alpha=e^{-\left(\lambda_1^2+...+\lambda_n^2\right)}$.\\

\noindent \textbf{Cavity-molecule spectroscopy} -- We set $N=1$ and consider laser driving of the cavity mode instead. The vibrational mode follows the same Langevin equation as in Eq.~\eqref{eq:Lang1} while the cavity-molecule coupling is described in terms of dressed polaron operators $\tilde{\s}(t)=\s (t)\Dd(t)$ and $\tilde{a}(t)=a (t)\Dd(t)$. We can formulate the dynamics in matrix form (see SM \cite{SM}) $\partial_t \bold{\tilde{v}}(t) = \textbf{M} \bold{\tilde{v}}(t)+ \bold{\tilde{v}}_{\text{in}}(t)$ for the vector of operators $\bold{\tilde{v}}(t)=(\tilde{\s}(t),\tilde{a}(t))^{\intercal}$. The dynamics is described by the drift matrix (with renormalized cavity frequency $\bomc =\omc -\sum_{\text{k}=1}^{n}\lambda_\text{k}^2\nuk$)
\begin{align}
\textbf{M} =
\begin{pmatrix}
-\left[\gamma-i(\omega_{\ell}-\ome)\right] & g\\
-g & -\left[\kappa-i(\omega_{\ell}-\bomc)\right]
\end{pmatrix},
\end{align}
and the vector of input noises contains the contributions from the displaced electronic and optical driving noises and is given by $\bold{\tilde{v}}_{\text{in}}(t)=(\sqrt{2\gamma}\Dd(t) \sigma_{\text{in}}(t),\sqrt{2\kappa}\Dd(t) A_{\text{in}}(t))^{\intercal}$ where we condensed here the input of the cavity in a single term $A_{\text{in}}=\eta_c/\sqrt{2\kappa}+a_{\text{in}}$. Diagonalization of the $\textbf{M}$ matrix implies that $\textbf{M}=\textbf{T} \Lambda_M \textbf{T}^{-1}$ where $\Lambda_M$ is a diagonal matrix with $\epsilon_{\pm}=-\left(\gamma_{\pm}+i\omega_{\pm}\right)$ on the diagonal. The $\textbf{T}$ matrix is made of the eigenvector components. A formal solution in steady state for the bare operator vector $\bold{v}(t)=\left(\s(t),a(t)\right)^{\intercal}$ can be obtained
\begin{align}
\bold{v}(t) = \int_{-\infty}^t dt' \textbf{T} e^{\Lambda_M (t-t')} \textbf{T}^{-1}\D(t)\bold{\tilde{v}}_{\text{in}}(t').
\end{align}
In general one is interested in deriving the cavity transmission function as the laser detuning is swept around the cavity resonance which we define as $\text{t}_\text{c} =\kappa \braket{a(t)}/\eta_{\text{c}}$. To this end we evaluate the second element of the operator vector and perform a classical averaging which leads to

\begin{align}
\text{t}_\text{c} &= \frac {\text{T}_{11}\text{T}_{22}}{\det \T}\sum_{{{\left\{m_\text{k}\right\}}}={0}}^{\infty} \frac{\kappa \text{s}_{m_1}^{\lambda_1}...\text{s}_{m_n}^{\lambda_n}}{\gamma_- + \Gamma_{\left\{m_\text{k}\right\}} -i\left(\oml-\omega_- - \nu_{\left\{m_\text{k}\right\}}\right)}\\ \nonumber
&-\frac {\text{T}_{12}\text{T}_{21}}{\det \T} \sum_{{{\left\{m_k\right\}}}={0}}^{\infty} \frac{\kappa \text{s}_{m_1}^{\lambda_1}...\text{s}_{m_n}^{\lambda_n}}{\gamma_+ + \Gamma_{\left\{m_\text{k}\right\}} -i\left(\oml-\omega_+ - \nu_{\left\{m_\text{k}\right\}}\right)}.
\label{cavityamplitude}
\end{align}
where ${\det \T} =\text{T}_{11}\text{T}_{22}-\text{T}_{12}\text{T}_{21}$. The poles of the denominator show that the maximum transmission occurs at $\omega_{\pm}$ with linewidths $\gamma_{\pm}$. For $\lambda=0$ (pure quantum emitter case) the strong coupling regime with $g \gg (\kappa,\gamma)$ predicts two eigenfrequencies roughly positioned at $\ome\pm g$ with linewidths $(\kappa+\gamma)/2$. For $\lambda>0$, dissipatively-induced polariton cross-talk leads to an asymmetric transmission profile as illustrated in Fig.~\ref{fig1}(d). A first order perturbative approach (as derived in SM \cite{SM}) including upper-lower polariton exchanges via single vibrations leads to an approximate dissipative rate:
\begin{align}
\kappa_{\text{UL}}= \frac{1}{2}\frac{\lambda^2\nu^2\Gamma}{\Gamma^2+\left(\omega_+-\omega_--\nu\right)^2}.
\end{align}
For $g=\nu/2$, the upper polariton lifetime can be simply casted as $2\kappa'=2\kappa[1+\lambda^2g^2/(\kappa\Gamma)]$. The upper polariton is a Lorentzian with an effectively reduced maximum corresponding to an increased linewidth $(\kappa+\gamma)/2+\kappa_{\text{UL}}/2$.\\
\noindent Another interesting aspect is the modification of the radiative branching ratio in the Purcell regime (experimentally employed to turn a molecule into an efficient quantum emitter~\cite{wang2019turning}). We analyze the fluorescence spectrum of a partially excited molecule for $\kappa\gg g_{\text{00}}\gg \gamma$ but large cooperativity $C_{\text{00}}=g_{\text{00}}^2/(\kappa \gamma)\gg 1$. We denote by $g_{\text{00}}=g e^{-\sum_{\text{k}=1}^{n}\lambda_\text{k}^2/2}$ the coupling rate between the cavity mode and the zero-phonon line reduced from the total rate $g$ by the reduction in oscillator strength. Assuming very good resolution of the zero-phonon line with $\omc=\ome$ and $\kappa\ll\nu_\text{k}$ the emergence of the Purcell regime is accompanied by $\omega_+=\omega_-=\ome$, $\gamma_+\approx\kappa$ but an enlarged $\gamma_-=\gamma'+\gamma_{\text{00}}(1+C_{\text{00}})\gg \gamma$. While emission into levels other than the zero-phonon line is unperturbed and sums up to $\gamma'=\gamma-\gamma_{\text{00}}$, the zero-phonon line decay is strongly enhanced and the  cavity-modifed branching ratio can be expressed as
\begin{align}
\alpha_{\text{cav}}=\frac{(1+C_{\text{00}})e^{-\sum_{\text{k}=1}^{n}\lambda_\text{k}^2}}{1+C_{\text{00}} e^{-\sum_{\text{k}=1}^{n}\lambda_\text{k}^2}},
\end{align}
going to unity in the limit that $C_{\text{00}}\rightarrow\infty$ and reproducing the bare molecule result for $C_{\text{00}}=0$.\\

\begin{figure}[t]
\includegraphics[width=0.92\columnwidth]{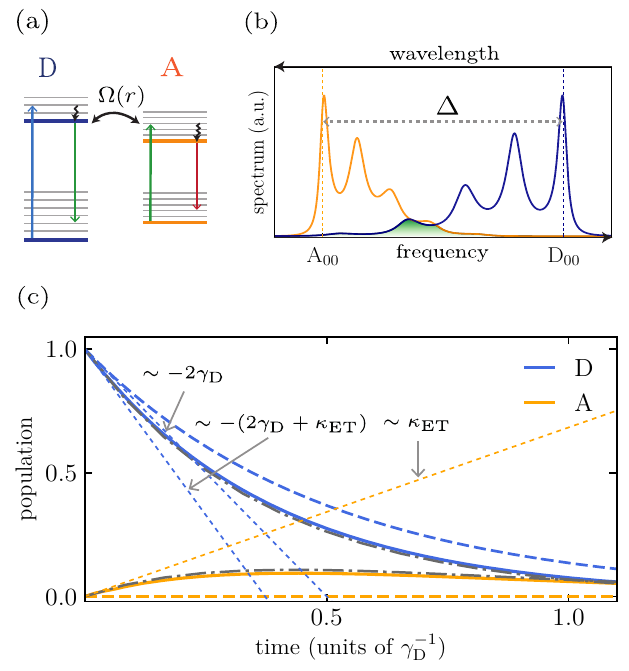}
\caption{(a) Unidirectional near-field FRET processes showing donor excitation (blue) followed by energy exchanges (green) and acceptor emission (red). (b) Donor emission and acceptor absorption spectra with overlap in green region. Zero phonon lines are indicated by $\text{A}_{\text{00}}$ and $\text{D}_{\text{00}}$. c) Solution of the rate equations for time dynamics of the D-A system with (full lines) and without coupling (dashed lines) in a pump-probe scenario. The tangents (dotted lines) show that the independent radiative emission of the donor at rate $2\gamma_{\text{D}}$ is modified by energy transfer rate $\kappa_{\text{ET}}$ while the initial population increase of the acceptor is governed by the same quantity $\kappa_{\text{ET}}$. Agreement with numerical simulations is indicated by the dashed-dotted curves with parameters $\gamma_A=\gamma_D=\gamma$, $\Omega(r)=15\gamma$,  $\Delta=500\gamma$, $\nu_A=\nu_D=250\gamma$, $\Gamma_A=\Gamma_D=30\gamma$, $\lambda_D=0.6$, $\lambda_A=0.4$.}
\label{fig2}
\end{figure}

\noindent \textbf{F\"{o}rster resonance energy transfer} -- We consider two molecules (donor-D and acceptor-A) with frequency mismatch $\omega_\text{D}-\omega_\text{A}=\Delta\gg\nu_\text{D,A}$ [see Fig.~\ref{fig2}(a)]. As Fig.~\ref{fig2}(b) illustrates, their emission-absorption spectra contain a small overlap which allows an energy transfer from the donor to the acceptor via near field strongly distance dependent dipole-dipole exchanges $\Omega(r)\propto 1/r^3$. For $\Omega(r)>\gamma_D$, an initially excited donor will lose energy both via radiative emission at rate $2\gamma_{\text{D}}$ as well as via a resonant exchange with the acceptor's higher excited state vibrational manifold. In a standard regime where $\Omega(r)\ll\Gamma_\text{A}$ a uni-directional process emerges as the acceptor quickly relaxes to the zero vibrational state and radiates around its peak emission frequency. We analyze the process in a transient pump-probe scenario where the donor is quickly excited and its natural decay rate is modified by a rate $\kappa_{\text{ET}}$ at which it spills energy into the acceptor via the emission-absorption overlap. We proceed by deriving effective rate equations: $\partial_t P_\text{D}=-(2\gamma_\text{D}+\kappa_{\text{ET}})P_\text{D}$ and $\partial_t P_\text{A}=-2\gamma_\text{A} P_\text{A}+\kappa_{\text{ET}}P_\text{D}$ for acceptor/donor excited state populations $P_\text{A/D}$ under the perturbation condition $\Omega(r)\ll\Gamma_\text{A}$. From the Langevin equations of motion (see SM \cite{SM}) we identify the term responsible for energy migration as $-2 \Omega(r)\Im\braket{\sd_\text{A}\s_\text{D}}\approx -\kappa_{\text{ET}}P_\text{D}(t)$ and estimate (for simplicity we set $\gamma_\text{D}=\gamma_\text{A}$) the F\"{o}rster resonance energy transfer as
\begin{align}
\kappa_{\text{ET}}=\sum_{n_\text{A},n_\text{D}}\frac{2\text{s}_{n_\text{D}}^{\lambda_\text{D}}\text{s}_{n_\text{A}}^{\lambda_\text{A}}(n_\text{A}\Gamma_\text{A}+n_\text{D}\Gamma_\text{D})\Omega(r)^2 }{(n_\text{A}\Gamma_\text{A}+n_\text{D}\Gamma_\text{D})^2+(\Delta-n_\text{A}\nu_\text{A}-n_\text{D}\nu_\text{D})^2}.
\end{align}
The scaling with distance is straightforward as $\Omega(r)^2 \propto r^{-6}$. Optimization of the transfer rate requires maximizing the Poissonian coefficients while bringing the Lorentzian into resonance. This is none other than the usual overlap integral widely discussed in literature but expressed in terms of discrete emission/absorption lines. The resonance transfer occurs when emission meets absorption i.e., fluorescence lines of the donor $\omega_\text{D}-n_\text{D}\nu_\text{D}$ overlap absorption lines of the acceptor $\omega_\text{A}+n_\text{A} \nu_\text{A}$. The formalism allows for extensions to many vibrational modes, temperature effects as well as for the inclusion of dissipative collective effects (subradiance/superradiance) which have been shown to play a role in enhancing the energy transfer rate between quantum emitters~\cite{reitz2018energy}. The pump-prope dynamics is illustrated in Fig.~\ref{fig2}(c) where the donor evolution sees an accelerated rate in the presence of a nearby placed acceptor (line tangent show the convergence of the early time evolution with the analytically estimated transfer rate). \\
In addition to the direct dipole-dipole interaction cavity-mediated processes via couplings $g_\text{D,A}$ between donor/acceptor and mode $a$ participate in the energy transfer. We focus on deriving scaling laws in the bad-cavity regime (large cooperativity) via the same perturbative method applied in the case of the direct F\"{o}rster energy transfer (see SM \cite{SM}): this leads to an extra term in the acceptor population rate equation $\partial_t P_\text{A}=-2\gamma_\text{A} P_\text{A}+\kappa_{\text{ET}}(r)P_\text{D}+\kappa_{\text{ET}}^\text{cav}P_\text{D}$. The first term contains a cavity-modified dipole-dipole energy transfer rate springing from $2\Omega(r)\Im\braket{\sd_\text{A}\s_\text{D}}$. The second term includes purely cavity mediated processes coming from $2g_A\Im\braket{\sd_\text{A}a}$. While the analytical expressions are cumbersome, an immediate result emerges which implies an acceleration of the transfer rate when the cavity is matched to the peak absorption line of the acceptor roughly at $\omega_c=\omega_\text{A}+n_\text{A}\nu_\text{A}$ with $n_\text{A}$ around $\lambda_{\text{A}}^2\nu_{\text{A}}$. Further optimization requires the particular details of the donor/acceptor emission/absorption lines and their relevant vibrational levels involved in the transfer.\\

\noindent \textbf{Conclusions -} A treatment based on quantum Langevin equations allows for analytical insight into the role of vibrations in cavity QED with molecules. Its strength lies in the inclusion of an arbitrary number of vibrations in thermal equilibrium at any T. We have derived spectroscopic quantities and presented general expressions for energy transfer rates between donor-acceptor molecules outside and inside cavities as well as a perturbative expression for dissipative polariton coupling. The approach can be extended to include the dynamics of bulk phonons and their coupling to vibrations leading to vibrational relaxation and phonon wings. Future endeavors will also aim at treating mesoscopic systems in the weak excitation limit with the input-output formalism of Ref.~\cite{plankensteiner2019enhanced} providing a quantum description of the cavity output field. This could allow for efficient numerical simulations of the collective strong coupling regime of cavity QED with molecules, identifying the role of dipole-dipole interactions, super/subradiant and collective dark states.

\vspace{0.1cm}
\noindent\textbf{Acknowledgments -} We acknowledge financial support from the Max Planck Society and from the German Federal Ministry of Education and Research, co-funded by the European Commission (project RouTe), project number 13N14839 within the research program "Photonik Forschung Deutschland" (C.~S. and C.~G.). We acknowledge fruitful discussions with T.~Ebbesen, C.~Genet, V.~Sandoghdar, S.~G\"{o}tzinger, D.~Martin-Cano and B. G\"{u}rlek.

\bibliography{bibfileQEDmoleculesMay}

\newpage
\onecolumngrid

\appendix

\section{Ad-hoc derivation of the Holstein model for a single nuclear coordinate}
\label{A}

Let us justify the form of the Holstein Hamiltonian \cite{holstein1959study} by following a first-principle derivation for a single nuclear coordinate $R$ of effective mass $\mu$. We assume that, along the nuclear coordinate, the equilibria for ground (coordinate $R_{\text{g}}$, state vector  $\ket{g}$) and excited (coordinate $R_{\text{e}}$ and state vector $\ket{e}$) electronic orbitals are different. Notice that, for the simplest case of a homonuclear diatomic molecule made of nuclei each with mass $m$, a single vibrational mode exists corresponding to the relative motion mode with effective mass $\mu = m/2$; the equilibria coordinates $R_{\text{g}}$ and $R_{\text{e}}$ correspond to the bond length in ground/excited states. Assuming that both the ground and the excited state have the same parabolic shape around the minima (with second derivative leading to a vibrational frequency $\nu$) one can write the total Hamiltonian:
\begin{equation}
H=\left[\omega_e+\frac{p^2}{2\mu}+\frac{1}{2}\mu\nu^2\left(R-R_e\right)^2\right]\sigma^\dagger\sigma + \left[\frac{p^2}{2\mu}+\frac{1}{2}\mu\nu^2\left(R-R_g\right)^2\right]\sigma\sigma^\dagger.
\end{equation}
Notice that the Hamiltonian is written in a Hilbert space spanning both electronic dynamics (via the Pauli operators) and mechanical dynamics. Introducing small oscillations around the equilibria $Q=R-R_{\text{g}}$ and subsequently $R-R_{\text{e}}=Q+R_{\text{g}}-R_{\text{e}}=:Q-R_{\text{ge}}$ we obtain
\begin{equation}
H=\frac{P^2}{2\mu}+\frac{1}{2}\mu\nu^2 Q^2 +\omega_e\sigma^\dagger\sigma-\mu\nu^2 Q R_{\text{ge}}\sigma^\dagger\sigma+\frac{1}{2}\mu\nu^2 R_{\text{ge}}^2 \sigma^\dagger\sigma.
\end{equation}
The last term is a renormalization of the bare electronic transition frequency energy (absent in some theoretical treatments) which will naturally go away when diagonalizing the Hamiltonian via the polaron transformation resulting in $\ome$ as the natural electronic transition. We can now rewrite the momentum and position operators in terms of bosonic operators $Q=r_{\text{zpm}}(b^\dagger+b)$, $P=ip_{\text{zpm}}(b^\dagger-b)$ by introducing the zero-point-motion $r_{\text{zpm}}=1/\sqrt{2\mu\nu}$ and $p_{\text{zpm}}=\sqrt{\mu\nu/2}$. Reexpressing the terms above via the Huang-Rhys factor $\lambda=\mu\nu R_{\text{ge}}r_{\text{zpm}}$ yields the Holstein-Hamiltonian
\begin{equation}
H=\nu \bd b+(\ome+\lambda^2\nu)\sd\s-\lambda\nu(\bd+b)\sd \s.
\label{holstein}
\end{equation}
One can bring this Hamiltonian into diagonal form $\nu \bd b+\omega_e\sd\s$ via the polaron transformation $U^{\dagger}=(\Dd)^{\sd\s}$ where the displacement is defined as $\D=\exp(-i\sqrt{2}\lambda p)=\exp[\lambda (\bd-b)]$ with the dimensionless momentum quadrature $p=i(\bd-b)/\sqrt{2}$ as generator (we also define the dimensionless position quadrature $q=(\bd+b)/\sqrt{2}$). Notice that the polaron transformation also removes the vibronic shift of the excited state.

\section{Quantum Langevin equations}
\label{B}

We will list the exact and approximated quantum Langevin equations for a single vibration with the observation that the many vibrations case is a straightforward generalization. The derivation is based on the property that an exact transformation exists linking any master equation in Lindblad form to an equivalent equation of motion that properly includes noise terms associated with the Lindblad collapse operators \cite{gardiner2004quantum}. Explicitly, for any operator $\cal{O}$
\begin{align}
\dot{\cal{O}}=i\left[H,\cal{O}\right]-\left[{\cal{O}},c^\dagger\right]\left\{\gamma_c c-\sqrt{2\gamma_c}c_{\mathrm{in}}\right\}+\left\{\gamma_c c^\dagger -\sqrt{2\gamma_c}c_{\mathrm{in}}^\dagger \right\}\left[{\cal{O}},c \right],
\end{align}
where an input noise occurs for any collapse operator $c$ at rate $\gamma_c$. In our case we have collapse operators $\s$ and $b$ with corresponding rates $\gamma$ and $\Gamma$. After working out the commutation relations, one obtains:
\begin{subequations}
\begin{align}
\frac{d b}{dt} &= -(\Gamma+i\nu)b +i\lambda\nu\sd\s+\sqrt{2\Gamma}\binp,\\
\frac{d\tilde{\s}}{dt} &= -(\gamma+i\ome)\tilde{\s} +\eta_{\ell} e^{-i\omega_{\ell}t} (1-2\sd\s)\Dd+\lambda \Gamma (\bd-b) \tilde{\s}+(1-2\sd\s)\sqrt{2\gamma}\tilde{\s}_{\mathrm{in}}-\sqrt{2\Gamma}\lambda\tilde{\s} (\binpd-\binp),\\
\frac{d \sd \s}{dt} &= -2\gamma \sd \s +\eta_{\ell}(\sd e^{-i\omega_{\ell}t}+\s e^{i\omega_{\ell}t})+\sqrt{2\gamma}\left[\sd \siginp+\siginpd \s\right].
\end{align}
\end{subequations}
Under the assumption of weak driving $\eta_\ell\ll\gamma$ the transition dipole scales at most as $\eta_\ell/\gamma$ while the population scales as $\eta_\ell^2/\gamma^2$. The back-action term leading to a change in the vibrational dynamics is then small compared to $\Gamma$ or $\nu$ which allows us to work in perturbation theory up to second order in $\eta_\ell$. We also can neglect terms $\sd\s$ with respect to unity. This leads to the following perturbative Langevin equations (where now we transform into a frame rotating at the laser frequency) which will be the basis of our analytical treatment:
\begin{subequations}
\label{LangSimpleAppendix}
\begin{align}
\frac{d b}{dt} &= -(\Gamma+i\nu)b +\sqrt{2\Gamma}\binp,\\
\label{analyticalequations}
\frac{d\tilde{\s}}{dt} &=  -(\gamma+i\ome-i\oml)\tilde{\s} +\eta_{\ell}\Dd(t) +\sqrt{2\gamma}\tilde{\s}_{\mathrm{in}},\\
\frac{d \sd \s}{dt} &= -2\gamma \sd \s +\eta_\ell(\sd +\s )+\sqrt{2\gamma}\left[\sd \siginp+\siginpd \s\right].
\end{align}
\end{subequations}
Here, in \eqref{analyticalequations} we have neglected the term $\lambda \Gamma (\bd-b) \tilde{\s}$ as well as the corresponding noise term $-\sqrt{2\Gamma}\lambda(b_{\text{in}}^\dagger-b_{\text{in}})\tilde{\s}$. We can however estimate the contribution of those terms by averaging (tracing) over the vibrational ground state $\ket{0}$ (assuming fast vibrational relaxation):
\begin{align}
\lambda \Gamma \braket{(\bd-b) \tilde{\s}}-\sqrt{2\Gamma}\lambda\braket{(b_{\text{in}}^\dagger-b_{\text{in}})\tilde{\s}}=-\lambda^2\Gamma\sigma e^{-\lambda^2/2},
\end{align}
which has to be compared with the bare electronic contribution $-\gamma\braket{\tilde{\sigma}}=-\gamma\sigma e^{-\lambda^2/2}$.
We can see that the terms effectively lead to an additional dephasing of the dipole operator with $\gamma+\lambda^2\Gamma$. This would imply disagreement with experimental observations of lifetime limited transitions for the zero-phonon line of molecules as vibrational relaxation rates are much larger than radiative rates. However, we show in (SM~\ref{Cprime}) that a modelling of non-unitary dynamics based on a Brownian noise model removes this inconsistency at the level of linewidths while showing perfect agreement in spectroscopic line positions. However, for simplicity of derivations and comparison with numerical simulations we rely on the simple Lindblad form master equation in the following.\\

The procedure we will follow to derive molecular spectra is to first integrate the vibrational dynamics $b(t)$ and subsequently replace the time evolved displacement operator $\D(t)$ in the equations for $\tilde{\s}(t)$ and $\sd \s(t)$. Most of the physics is contained in the two-time correlations $\braket{\D(t)\Dd(t')}$ which we evaluate in Appendix~\ref{C}.

\section{Dynamics of the vibrational mode. Properties of the displacement operators}
\label{C}
We integrate here the equations of motion for the vibrational operators and compute the two-time correlation functions for displacement operators as relevant to the further computation of any electronic degree of freedom operator.

\subsection*{Simplified model: creation and annihilation operators.}
First, let us analyze the properties of the vibrational mode operators. We will follow the dynamics on timescales defined by the slow radiative evolution with a timescale $\gamma^{-1}\gg\Gamma^{-1}$. This allows us to consider that the vibrations are practically always in steady state and operators at any time $t>0$ are given by a formal integration with initial conditions extended to $-\infty$ such that
\begin{align}
b(t) =\sqrt{2\Gamma} \int_{-\infty}^t d\tau e^{-(\Gamma+i\nu)(t-\tau)} \binp(\tau).
\end{align}
The corresponding Hermitian conjugate operator is given by
\begin{align}
\bd(t) =\sqrt{2\Gamma} \int_{-\infty}^t d\tau e^{-(\Gamma-i\nu)(t-\tau)} \binpd(\tau),
\end{align}
and the commutator of such operators results in
\begin{align}
[b(t), \bd(t')] = \left\{\begin{array}{cc} e^{-(\Gamma+i\nu)(t-t')} & t\geq t' \\ e^{-(\Gamma-i\nu)(t'-t)} & t'\geq t \end{array} \right. .
\end{align}

\subsection*{Spectrum of the Holstein force. Non-zero temperature bath.}

Let us estimate the power spectrum of the force acting on the excited electronic orbital by defining (in analogy with Ref.~\cite{neuman2018origin}) an effective hermitian force $F=\lambda \nu (b+\bd)$. The power spectrum is defined as:
\begin{align}
F(\omega)= 2\Re \left\{\int_{0}^{\infty} d\tau e^{i\omega \tau}\braket {F(\tau)F(0)}\right\}.
\end{align}
which is none other than the spectrum of the displacement variable in equilibrium with the environment. Notice that the only non-vanishing correlations at different times are:
\begin{align}
\braket{b(t)\bd(t')} =2 \Gamma \int_{-\infty}^t d\tau \int_{-\infty}^{t'}  d\tau' e^{-(\Gamma+i\nu)(t-\tau)}e^{-(\Gamma-i\nu)(t'-\tau')} \delta(\tau-\tau')=e^{-(\Gamma+i\nu)(t-t')},
\end{align}
which leads to a spectrum of the effective force acting on the excited electronic orbital of the molecule
\begin{align}
F(\omega)= 2\lambda^2 \nu^2 \Re \left\{\frac{1}{\Gamma+i(\omega-\nu)}\right\} = \frac{2\lambda^2 \nu^2 \Gamma}{\Gamma^2+(\omega-\nu)^2}.
\end{align}
One can also include a $T>0$ bath in the correlations of the noise operators $\braket{b_{\text{in}}^\dagger(t)b_{\text{in}}(t')}=\Gamma \bar{n} \delta(t-t')$ and $\braket{b_{\text{in}}(t)b_{\text{in}}^\dagger(t')}=\Gamma (\bar{n}+1)\delta(t-t')$ where the thermal occupancy $\bar{n}=[\exp({\hbar \nu/(k_\text{B}\text{T})})-1]^{-1}$. This would lead to a temperature dependent spectrum
\begin{align}
F(\omega;T)= 2\lambda^2 \nu^2 \Re \left\{\frac{1}{\Gamma+i(\omega-\nu)}\right\}(1+2\bar{n}) = \frac{2\lambda^2 \nu^2 \Gamma}{\Gamma^2+(\omega-\nu)^2}(1+2\bar{n}).
\end{align}

\subsection*{Two-time correlations of displacement operators.}

Now we can estimate properties of the displacement operator. The trace over the displacement operators in the Heisenberg picture is done with the initial vacuum density operator $\ket{0}\bra{0}$ for the vibrational mode (as well as for all the outside modes encompassed in the input noise). We can then deduce the two-time correlations for a given time ordering $t>t'$
\begin{equation}
\begin{split}
\braket{0|\mathcal{D}(t)\mathcal{D}^\dagger(t')|0}&= \braket{0|e^{\lambda(b^\dagger (t)-b(t))}e^{-\lambda(b^\dagger (t')-b(t'))}|0} = \braket{0|e^{-\lambda b (t)} e^{-\lambda b^\dagger (t')}|0}e^{-\lambda^2}= e^{\lambda^2\left[e^{-\left(\Gamma+i\nu\right)\left(t-t'\right)}\right]}e^{-\lambda^2}.
\end{split}
\end{equation}
This leads to the main expressions that we will be using in the following especially in the sum form expressed below:
\begin{align}
\braket{\D(t)\Dd(t')}=e^{-\lambda^2} e^{\lambda^2 \left[e^{-(\Gamma+i\nu)(t-t')}\right]} =\sum_{m=0}^{\infty} s_{m}^{\lambda} e^{-m(\Gamma+i\nu)(t-t')},
\end{align}
and
\begin{align}
\braket{\D(t')\Dd(t)}=e^{-\lambda^2} e^{\lambda^2 \left[e^{-(\Gamma-i\nu)(t-t')}\right]} =\sum_{m=0}^{\infty} s_{m}^{\lambda} e^{-m(\Gamma-i\nu)(t-t')},
\end{align}
where $s^{\lambda}_{m} = e^{-\lambda^2}\lambda^{2m}/m!$.

\section{Brownian noise model.}
\label{Cprime}

The dynamics of a mechanical oscillator subject to thermal noise (modeled as Brownian motion dynamics) is encompassed in the following equations of motion for the dimensionless quadratures $q=(b^\dagger+b)/\sqrt{2}$ and $p=i(b^\dagger-b)/\sqrt{2}$
\begin{align}
\dot{q}(t) &= \nu p,\\
\dot{p}(t) &= -\nu q-2\Gamma p + \zeta(t),
\end{align}
combined with the correlations of the input noise at a given bath temperature in the time domain \cite{genes2008ground}
\begin{align}
\braket{\zeta(t)\zeta(t')}=\frac{1}{2\pi}\int_{-\infty}^{\infty}d\omega e^{-i\omega(t-t')}\left\{\frac{2\Gamma\omega}{\nu}\left[\coth{\frac{\hbar \omega}{2k_\text{B} T}}+1\right]\right\}=\frac{1}{2\pi}\int_{-\infty}^{\infty}d\omega e^{-i\omega(t-t')}S_\text{th}(\omega).
\end{align}
For an arbitrary temperature, the average thermal occupancy is $\bar{n}=1/(\exp(\hbar \omega/k_\text{B} T)-1)$. The spectrum at the $\pm \nu$ then gives the cooling and heating contributions: $S_\text{th}(\nu)=4\Gamma(\bar{n}+1)$ and $S_\text{th}(-\nu)=4\Gamma\bar{n}$. Generally, from these properties one can always estimate the damping rate $2\Gamma=(S_\text{th}(\nu)-S_\text{th}(-\nu))/2$ and equilibrium occupancy as $\bar{n}=S_\text{th}(-\nu)/[S_\text{th}(\nu)+S_\text{th}(-\nu)]$. Close to zero temperature $T=0$, the thermal spectrum becomes very asymmetric $S_\text{th}(\omega)=\left[4\omega \Gamma/\nu\right] \theta(\omega)$ as it vanishes for negative frequencies (we introduced the Heaviside function $\theta(\omega)$). Generally, one can proceed with a Fourier domain analysis of the steady state of the equations above which lead to:
\begin{align}
q(\omega)=\epsilon(\omega)\zeta(\omega) \qquad \text{and} \qquad p(\omega)=-i\frac{\omega}{\nu}\epsilon(\omega) \zeta(\omega) \qquad \text{and} \qquad \epsilon(\omega)=\frac{\nu}{\nu^2-\omega^2-2i\Gamma \omega}.
\end{align}
The zero point energy can be estimated by using $\int_{0}^{\infty}d\omega|\epsilon(\omega)|^2=\pi/(4\Gamma)$ which leads to $\braket{q^2}=1/2$ as expected. However, for the momentum the integral $\int_{0}^{\infty}d\omega|\omega\epsilon(\omega)|^2$ diverges and requires the introduction of a cutoff frequency $\Lambda>\nu$ to keep it finite. Under this assumption and noticing that the susceptibility is sharply peaked at $\nu$ one can evaluate the integral by fixing $\omega=\nu$ in the numerator and obtaining the expected equipartition rule. Furthermore, one can obtain the following two properties
\begin{align}
\left[ q(t),p(t')\right] & = i\left\{\begin{array}{cc} e^{-(\Gamma+i\nu)(t-t')} & t\geq t' \\ e^{-(\Gamma-i\nu)(t'-t)} & t'\geq t \end{array} \right. \\
\braket{p(t)p(t')}& = \frac{1}{2}\left\{\begin{array}{cc} e^{-(\Gamma+i\nu)(t-t')} & t\geq t' \\ e^{-(\Gamma-i\nu)(t'-t)} & t'\geq t \end{array} \right.,
\end{align}
which is identical to the Langevin case (when replacing $p=i(b^\dagger-b)/\sqrt{2}$, $q=(b^\dagger+b)/\sqrt{2}$). This also implies that the correlation function for the displacement operator $\braket{\D(t)\Dd(t')}$ is identical in both models.
Calculating the equation of motion for $\dot{\tilde{\sigma}}=\dot{\s}\mathcal{D}^\dagger+\s\dot{\mathcal{D}}^\dagger$ one finds
\begin{equation}
\dot{\tilde{\s}}= -\left[\gamma-i(\omega_\ell-\omega_e)\right]\tilde{\s} +\eta\Dd-2\sqrt{2}i\Gamma\lambda\sigma p \mathcal{D}^\dagger+\sqrt{2\gamma}\tilde{\s}_{\mathrm{in}}+\sqrt{2}i\sigma\lambda\zeta\mathcal{D}^\dagger.
\end{equation}
We can again estimate the contribution of the off-diagonal terms which gave rise to a dephasing term in the Langevin model. Here we obtain
\begin{align}
-2\sqrt{2}i\Gamma\lambda\sigma\braket{ p \mathcal{D}^\dagger}+\sqrt{2}i\sigma\lambda\braket{\zeta\mathcal{D}^\dagger}=0,
\end{align}
where we used that $\braket{p\mathcal{D}^\dagger}=\frac{1}{i\sqrt{2}}\frac{d}{d\lambda}\braket{\mathcal{D}^\dagger}$ and $\braket{\zeta\mathcal{D}^\dagger}=\sum_{m=0}^\infty\frac{(i\sqrt{2}\lambda)^m}{m!}\braket{\zeta(t)p(t)^m}=\sqrt{2}i\Gamma\lambda e^{-\lambda^2/2}$ (obtained by an application of the Isserlis' theorem and with relation $\braket{\zeta(t)p(t)}=\Gamma$).

\section{Absorption profile (steady state analysis)}
\label{D}
We will analyze the absorption profile of the system under continuous (weak) driving in steady state. This amounts to computing the expression of the excited state population $\braket{\sd \s}$ as a function of the incoming laser frequency.

\subsection*{Single vibrational mode.}

From Eqs.~\eqref{LangSimpleAppendix} we find:
\begin{align}
\frac{d}{dt}\braket{\sd \s} = -2\gamma \braket{\sd \s} +\eta_\ell(\braket{\sd} +\braket{\s})+\sqrt{2\gamma}\braket{\sd \siginp+\siginpd\s}.
\end{align}
To proceed we will need the expression of the Pauli operator in steady state which we obtain as:
\begin{align}
\s(t) = \int_{-\infty}^t dt' e^{-[\gamma-i(\oml-\ome)](t-t')}\left(\eta_\ell + \sqrt{2\gamma} \s_{\text{in}}\right)\D(t)\Dd(t').
\end{align}
To find the average molecular dipole one then performs a trace over the right hand side of the operator expression:
\begin{align}
\braket{\s(t)} = \eta_\ell \int_{-\infty}^t dt' e^{-[\gamma-i(\oml-\ome)](t-t')} \braket{\D(t)\Dd(t')}=\eta_\ell \sum_{m=0}^\infty \frac{s_m^\lambda}{(\gamma+m\Gamma)-i(\oml-\ome-m\nu)}.
\end{align}
Now we can first see that the last term is vanishing while the general expression can be evaluated as
\begin{align}
\braket{\sd(t)\s(t)} &= \int_{-\infty}^t dt' e^{-2 \gamma(t-t')} \eta_\ell\left[ \braket{\sd(t')} +\braket{\s(t')}\right]=\eta_\ell^2 \sum_{m=0}^\infty \frac{s_m^\lambda(\gamma+m\Gamma)/\gamma}{(\gamma+m\Gamma)^2+(\oml-\ome-m\nu)^2}.
\end{align}

\subsection*{Many vibrational modes.}

Extending to many modes we have the same equation as before
\begin{align}
\frac{d\tilde{\s}}{dt} = -\left[\gamma-i(\oml-\ome)\right]\tilde{\s} +\eta \Dd (t) + \sqrt{2\gamma} \tilde{\s}_{\text{in}},
\end{align}
except that now the displacement is produced by all vibrations
\begin{align}
 \Dd = \prod_{\text{k}=1}^{n}\D_{\text{k}}^\dagger=\D_1^\dagger\D_2^\dagger...\D_{n}^\dagger.
\end{align}
For the calculation of absorption and emission profiles we have to evaluate terms like
\begin{align}
& \int_{-\infty}^t dt' e^{-\left[\gamma-i(\oml-\ome)\right](t-t')} \braket{\D(t)\Dd(t')} = \int_{-\infty}^t dt' e^{-\left[\gamma-i(\oml-\ome)\right](t-t')} \braket{\D_1(t)\D_2(t)...\D_{n}(t) \D_1^\dagger(t')\D_2^\dagger(t')...\D_{n}^\dagger(t')}.
\end{align}
The expectation value can be factorized as modes are independent from each other at any given time and we use for each mode
\begin{align}
\braket{\D_{\text{k}}(t)\D_{\text{k}}^\dagger(t')} = e^{-\lambda^2_k} e^{\lambda^2_k \left[e^{-(\Gamma_k+i\nu_k)(t-t')}\right]} = \sum_{m_k=0}^{\infty} s_{m_k}^{\lambda_k} e^{-m_k(\Gamma_k+i\nu_k)(t-t')},
\end{align}
so we have to compute:
\begin{align}
 e^{-\sum_{k=1}^{n}\lambda^2_k} \int_{-\infty}^t dt' e^{-(\gamma-i(\omega_l -\omega_{\text{e}}))(t-t')}\left[e^{\lambda^2_1\left[e^{-(\Gamma_1+i\nu_1)(t-t')}\right]}..e^{\lambda^2_{n}\left[e^{-(\Gamma_{n}+i\nu_{n})(t-t')}\right]}\right].
\end{align}

\noindent We can now perform the integral to find

\begin{align}
\braket{\s(t)}=  \eta_\ell \sum_{\{m_{k}\}=0}^{\infty}  \frac{s^{\lambda_1}_{m_1}\cdots s^{\lambda_{n}}_{m_{n}}}{\gamma + \Gamma_{\{m_{k}\}} -i\left((\omega_l - \omega_{\text{e}}) - \nu_{\{m_{k}\}}\right)},
\end{align}
where $\Gamma_{\{m_{k}\}} = \sum_{k=1}^{n}m_{k}\Gamma_k$, $\nu_{\{m_{k}\}} = \sum_{k=1}^{n}m_{k}\nu_k$ and $\left\{m_k\right\}\equiv{m_1,...m_n}$ is a list of all indices. This leads to the corresponding population expression
\begin{align}
\label{pop}
\braket{\sd(t)\s(t)} &= \int_{-\infty}^t dt' e^{-2 \gamma(t-t')} \eta_\ell\left[ \braket{\sd(t')} +\braket{\s(t')}\right]=\eta_\ell^2 \sum_{\{m_k\}=0}^\infty \frac{s^{\lambda_1}_{m_1}\cdots s^{\lambda_{n}}_{m_{n}}(\gamma+\Gamma_{\{m_k\}})/\gamma}{(\gamma+\Gamma_{\{m_k\}})^2+(\oml-\ome-\nu_{\{m_k\}})^2}.
\end{align}

\section{Emission spectrum}
\label{E}

\subsection*{Transient regime.}
A simple way to obtain the emission spectrum \cite{auffeves2008spontaneous} is to consider the transient dynamics of a molecule initially partially excited with $\braket{\sigma^\dagger(0)\sigma(0)}=p_0$. From Eq.~\eqref{pauli} we can compute two-time correlations and eliminate all the vanishing ones (between transient and steady state contributions) to arrive at
\begin{align}
\braket{\sigma^\dagger(\tau)\sigma(0)}=p_0\braket{\mathcal{D}(0)\mathcal{D}^\dagger(\tau)}  e^{-\left(\gamma+i\ome\right)\tau}=\sum_m s_m^\lambda p_0 e^{-\left(\gamma+m\Gamma+i\left(\ome-m\nu\right)\right)\tau}.
\end{align}
Performing the Fourier transform gives
\begin{equation}
S_\text{{em}}(\omega)=2\Re\int_{0}^\infty d\tau e^{-i\omega\tau}\braket{\sigma^\dagger(\tau)\sigma(0)}= \sum_m s_m^\lambda \frac{2\left(\gamma+m\Gamma\right)p_0}{\left(\gamma+m\Gamma\right)^2+\left(\omega-\ome+m\nu\right)^2}.
\end{equation}

\subsection*{Steady state regime.}

In steady state the calculation of the emission spectrum is more cumbersome as it involves the evaluation of four-point correlation functions. Using the solutions
\begin{align}
\s(t) &= \int_{-\infty}^t d\tau e^{-(\gamma-i\Delta_e)(t-\tau)} \left(\eta_\ell + \sqrt{2\gamma} \s_{\text{in}}\right)\D(t)\Dd(\tau) \\
\sd(t) &= \int_{-\infty}^t d\tau e^{-(\gamma+i\Delta_e)(t-\tau)} \left(\eta_\ell + \sqrt{2\gamma} \sd_{\text{in}}\right)\D(\tau)\Dd(t),
\end{align}
in the steady state regime the autocorrelation function is given by
\begin{eqnarray}
\langle \sd(t) \s(t')\rangle &=& \eta_\ell^2\int^{t}_{-\infty}d\tau\int^{t'}_{-\infty}d\tau' e^{-(\gamma+i\Delta_{\text{e}})(t-\tau)}e^{-(\gamma-i\Delta_{\text{e}})(t'-\tau')}\langle \D(\tau)\Dd(t)\D(t')\Dd(\tau') \rangle.
\end{eqnarray}
The four-point correlation function of the displacement operators can be easily evaluated to be
\begin{eqnarray}
\langle \D(\tau)\Dd(t)\D(t')\Dd(\tau') \rangle &=& \langle 0|e^{\lambda(\bd(\tau)-b(\tau))}e^{-\lambda(\bd(t)-b(t))}e^{\lambda(\bd(t')-b(t'))}e^{-\lambda(\bd(\tau')-b(\tau'))} |0 \rangle \\\nonumber
&=& \langle 0|e^{-\lambda b(\tau)}e^{-\lambda \bd(t)}e^{\lambda b(t)}e^{\lambda \bd(t')}e^{-\lambda b(t')}e^{-\lambda \bd(\tau')} |0 \rangle e^{-2\lambda^2} \\\nonumber
&=& \langle 0|e^{-\lambda b(\tau)}e^{\lambda b(t)}e^{\lambda \bd(t')}e^{-\lambda \bd(\tau')} |0 \rangle e^{\lambda^{2}[e^{-(\Gamma+i\nu)(t'-\tau')}]}e^{\lambda^{2}[e^{-(\Gamma-i\nu)(t-\tau)}]}e^{-2\lambda^2}.
\end{eqnarray}
Here we have used the conditions $t \geq \tau$ and $t' \geq \tau'$.
The final derivations involve time ordering.
For an ordering given by $(1)$ $t > \tau > t' > \tau'$ we obtain the expression
\begin{eqnarray}
\label{Eq.6}
\langle \D(\tau)\Dd(t)\D(t')\Dd(\tau') \rangle_{(1)} &=& e^{\lambda^{2}[e^{-(\Gamma+i\nu)(t-t')}]}e^{-\lambda^{2}[e^{-(\Gamma+i\nu)(\tau-t')}]}e^{-\lambda^{2}[e^{-(\Gamma+i\nu)(t-\tau')}]} \\\nonumber
& & \times e^{\lambda^{2}[e^{-(\Gamma+i\nu)(\tau-\tau')}]}e^{\lambda^{2}[e^{-(\Gamma+i\nu)(t'-\tau')}]}e^{\lambda^{2}[e^{-(\Gamma-i\nu)(t-\tau)}]}e^{-2\lambda^2}.
\end{eqnarray}
Similarly, we obtain
\begin{eqnarray}
\label{Eq.7}
\langle \D(\tau)\Dd(t)\D(t')\Dd(\tau') \rangle_{(2)} &=& e^{\lambda^{2}[e^{-(\Gamma+i\nu)(t-t')}]}e^{-\lambda^{2}[e^{-(\Gamma-i\nu)(t'-\tau)}]}e^{-\lambda^{2}[e^{-(\Gamma+i\nu)(t-\tau')}]} \\\nonumber
& & \times e^{\lambda^{2}[e^{-(\Gamma+i\nu)(\tau-\tau')}]}e^{\lambda^{2}[e^{-(\Gamma+i\nu)(t'-\tau')}]}e^{\lambda^{2}[e^{-(\Gamma-i\nu)(t-\tau)}]}e^{-2\lambda^2},
\end{eqnarray}
for $(2)$ $t > t' > \tau > \tau'$ and
\begin{eqnarray}
\label{Eq.8}
\langle \D(\tau)\Dd(t)\D(t')\Dd(\tau') \rangle_{(3)} &=& e^{\lambda^{2}[e^{-(\Gamma+i\nu)(t-t')}]}e^{-\lambda^{2}[e^{-(\Gamma-i\nu)(t'-\tau)}]}e^{-\lambda^{2}[e^{-(\Gamma+i\nu)(t-\tau')}]} \\\nonumber
& & \times e^{\lambda^{2}[e^{-(\Gamma-i\nu)(\tau'-\tau)}]}e^{\lambda^{2}[e^{-(\Gamma+i\nu)(t'-\tau')}]}e^{\lambda^{2}[e^{-(\Gamma-i\nu)(t-\tau)}]}e^{-2\lambda^2}.
\end{eqnarray}
for $(3)$ $t > t' > \tau' > \tau$.
Injecting these sequences into the autocorrelation function we obtain the general result
\begin{eqnarray}
\label{Eq.9}
\nonumber
\langle \sd(t) \s(t')\rangle &=& \eta_\ell^2e^{-(\gamma+i\Delta_{\text{e}})t}e^{-(\gamma-i\Delta_{\text{e}})t'}\left(\int^{t}_{t'}d\tau\int^{t'}_{-\infty}d\tau' e^{(\gamma+i\Delta_{\text{e}})\tau}e^{(\gamma-i\Delta_{\text{e}})\tau'}\langle \D(\tau)\Dd(t)\D(t')\Dd(\tau') \rangle_{(1)}\right.\\\nonumber
& & + \int^{t'}_{-\infty}d\tau\int^{\tau}_{-\infty}d\tau' e^{(\gamma+i\Delta_{\text{e}})\tau}e^{(\gamma-i\Delta_{\text{e}})\tau'}\langle \D(\tau)\Dd(t)\D(t')\Dd(\tau') \rangle_{(2)} \\
& & \left. + \int^{t'}_{-\infty}d\tau'\int^{\tau'}_{-\infty}d\tau e^{(\gamma+i\Delta_{\text{e}})\tau}e^{(\gamma-i\Delta_{\text{e}})\tau'}\langle \D(\tau)\Dd(t)\D(t')\Dd(\tau') \rangle_{(3)}\right)\\\nonumber
&=& \eta_\ell^2\sum_{n_2,\dots , n_7}e^{4\lambda^2}s^{\lambda}_{n_2}s^{\lambda}_{n_3}s^{\lambda}_{n_4}s^{\lambda}_{n_5}s^{\lambda}_{n_6}s^{\lambda}_{n_7}(-1)^{n_{3}+n_{4}}\\\nonumber
& & \times \left[\frac{e^{-[(n_2 + n_3 + n_4 + n_5)(\Gamma+i\nu)](t-t')}-e^{-[\gamma + (n_2 + n_4 + n_7)\Gamma + i(\Delta_{\text{e}}+ (n_2+n_4-n_7)\nu)](t-t')}}{{\scriptstyle [\gamma+(n_7-n_3-n_5)\Gamma+i(\Delta_{\text{e}}-(n_3+n_5+n_7)\nu)][\gamma+(n_4+n_5+n_6)\Gamma-i(\Delta_{\text{e}}-(n_4+n_5+n_6)\nu)]}} \right. \\\nonumber
& & \left. + \frac{e^{-[\gamma + (n_2 + n_4 + n_7)\Gamma + i(\Delta_{\text{e}}+ (n_2+n_4-n_7)\nu)](t-t')}}{{\scriptstyle [2\gamma+(n_3+n_4+n_6+n_7)\Gamma+i(n_4+n_6-(n_3+n_7))\nu][\gamma+(n_4+n_5+n_6)\Gamma-i(\Delta_{\text{e}}-(n_4+n_5+n_6)\nu)]}} \right. \\\nonumber
& & \left. + \frac{e^{-[\gamma + (n_2 + n_4 + n_7)\Gamma + i(\Delta_{\text{e}}+ (n_2+n_4-n_7)\nu)](t-t')}}{{\scriptstyle [2\gamma+(n_3+n_4+n_6+n_7)\Gamma+i(n_4+n_6-(n_3+n_7))\nu][\gamma+(n_3+n_5+n_7)\Gamma+i(\Delta_{\text{e}}-(n_3+n_5+n_7)\nu)]}} \right].
\end{eqnarray}
For $t-t' = 0$ we obtain the result in Eq.~\ref{pop} for the evolution of the population $\braket{\sd(t)\s(t)}$. For $t-t' \gg 1/\Gamma$ the saturation value
\begin{align}
\braket{\sd(t)\s(t')} &= \eta_\ell^2\sum_{n_6=0}^{\infty}\frac{e^{-\lambda^2}s^{\lambda}_{n_6}}{[\gamma+i(\omega_l-\omega_{\text{e}})][\gamma + n_6\Gamma-i(\omega_l-\omega_{\text{e}}-n_6\nu)]}
\end{align}
is reached.

To calculate the emission spectrum we need to evaluate the Fourier transform of the autocorrelation function
\begin{eqnarray}
\label{Eq.Y}
S_{\text{em}}(\omega) &=& \int^{\infty}_{-\infty}d\tau\langle \sd(\tau)\s(0)\rangle e^{-i\omega\tau} = 2\Re\left[\int^{\infty}_{0}d\tau\langle \sd(\tau)\s(0)\rangle e^{-i\omega\tau}\right]\\
&=& 2\eta_\ell^{2}e^{4\lambda^{2}}\sum_{n_2,\dots,n_7}s^{\lambda}_{n_2}\dots s^{\lambda}_{n_7}(-1)^{n_3+n_4}\Re\left[ \frac{A_{n_{2}\dots n_7}(\Delta_{\text{e}})}{ [i\omega + (n_2+n_3+n_4+n_5)(\gamma_{\mathrm{vr}}+i\nu)]}\right. \\\nonumber
& & \left. + \frac{B_{n_2\dots n_7}(\Delta_{\text{e}})}{[\gamma + (n_2+n_4+n_7)\gamma_{\mathrm{vr}}+i(\omega+\Delta_{\text{e}} + (n_2 + n_4 - n_7)\nu)]}\right],
\end{eqnarray}
where
\begin{eqnarray}
\label{Eq.20}
A_{n_2\dots n_7}(\Delta_{\text{e}}) &=& \left( \frac{-1}{{\scriptstyle [\gamma+(n_7-n_3-n_5)\Gamma+i(\Delta_{\text{e}}-(n_3+n_5+n_7)\nu)][\gamma+(n_4+n_5+n_6)\Gamma-i(\Delta_{\text{e}}-(n_4+n_5+n_6)\nu)]}}\right)\\
B_{n_2\dots n_7}(\Delta_{\text{e}}) &=& \left( \frac{1}{{\scriptstyle [\gamma+(n_7-n_3-n_5)\Gamma+i(\Delta_{\text{e}}-(n_3+n_5+n_7)\nu)][\gamma+(n_4+n_5+n_6)\Gamma-i(\Delta_{\text{e}}-(n_4+n_5+n_6)\nu)]}} \right. \\\nonumber
& & -\frac{1}{{\scriptstyle [2\gamma + (n_3+n_4+n_6+n_7)\Gamma + i(n_4+n_6-(n_3+n_6))\nu]}}\\\nonumber
& & \left. \times \frac{{\scriptstyle [2\gamma + (n_3+n_4+n_6+n_7+2n_5)\Gamma + i(n_4+n_6-(n_3+n_6))\nu]}}{{\scriptstyle [\gamma+(n_3+n_5+n_7)\Gamma+i(\Delta_{\text{e}}-(n_3+n_5+n_7)\nu)][\gamma+(n_4+n_5+n_6)\Gamma-i(\Delta_{\text{e}}-(n_4+n_5+n_6)\nu)]}}\right).
\end{eqnarray}

\section{Molecule-cavity spectroscopy}
\label{F}

\subsection*{Langevin equations.}

The master equation to Langevin equation procedure is now applied to the dynamics including the Jaynes-Cummings interaction and decay at rate $\kappa$ of the cavity mode. Notice that the natural basis is that of polaron operators both for the field mode $\tilde{a}=a\D^\dagger$ and molecule $\tilde{\s}=\s \D^\dagger$:
\begin{subequations}
\begin{align}
\frac{d b}{dt} &= -(\Gamma+i\nu)b +i\lambda\nu\sd\s+\sqrt{2\Gamma}\binp,\\
\frac{d\tilde{\s}}{dt} &= -(\gamma+i\ome)\tilde{\s} -g \sigma_z \tilde{a}+\lambda \Gamma (\bd-b) \tilde{\s}+(1-2\sd\s)\sqrt{2\gamma}\tilde{\s}_{in}-\lambda\sqrt{2\Gamma} (\binpd-\binp)\tilde{\s},\\
\frac{d\tilde{a}}{dt} &= -\left[\kappa+i\left(\omc-\lambda^2 \nu\right)\right]\tilde{a}-g \tilde{\s}+\lambda \Gamma (\bd-b) \tilde{a}+\eta_c e^{-i\omega_{\ell}t}\Dd+\sqrt{2\kappa} \tilde{a}_{in}-\lambda\sqrt{2\Gamma}(\binpd-\binp)\tilde{a}.
\end{align}
\end{subequations}
With the simplifications assumed before (weak excitation) the following set of solvable equations emerges:
\begin{subequations}
\begin{align}
\frac{d b}{dt} &= -(\Gamma+i\nu)b +\sqrt{2\Gamma}\binp,\\
\frac{d\tilde{\s}}{dt} &= -(\gamma+i\ome)\tilde{\s}+ g \tilde{a}+\sqrt{2\gamma}\tilde{\s}_{\mathrm{in}},\\
\frac{d\tilde{a}}{dt} &= -\left[\kappa+i\left(\omc-\lambda^2 \nu\right)\right]\tilde{a} -g \tilde{\s}+\eta_c e^{-i\omega_{\ell}t}\Dd+\sqrt{2\kappa} \tilde{a}_{\mathrm{in}}.
\end{align}
\end{subequations}

\subsection*{Cavity transmission.}

While the first equation is independent, the last two are coupled and we write them in matrix form:
\begin{align}
\frac{d\bold{\tilde{v}}}{dt} = \textbf{M} \bold{\tilde{v}}+ \bold{\tilde{v}}_{\mathrm{in}},
\end{align}
where the drift matrix is
\begin{align}
\textbf{M} =
\begin{pmatrix}
-\left[\gamma+i(\ome-\omega_{\ell})\right] & g\\
-g & -\left[\kappa+i(\omc-\lambda^2 \nu-\omega_{\ell})\right]
\end{pmatrix},
\end{align}
accompanied by a transformation into a frame rotating with the laser frequency.
The input vector is
\begin{align}
\bold{\tilde{v}}_{\mathrm{in}} =
\begin{pmatrix}
\sqrt{2\gamma} \tilde{\s}_{\mathrm{in}} \\
\eta_c \Dd +\sqrt{2\kappa} \tilde{a}_{\mathrm{in}}
\end{pmatrix}
=\begin{pmatrix}
\sqrt{2\gamma} \s_{\mathrm{in}} \\
\eta_c +\sqrt{2\kappa} a_{\mathrm{in}}
\end{pmatrix}
\Dd(t).
\end{align}
Diagonalization of the $\textbf{M}$ matrix implies that $\textbf{M}=\textbf{T} \boldsymbol{\Lambda}_M \textbf{T}^{-1}$ where $\boldsymbol{\Lambda}_M$ is a diagonal matrix with $\epsilon_{\pm}$ on the diagonal. The $\textbf{T}$ matrix is made of the eigenvector components. We can now write
\begin{align}
\textbf{T}^{-1}\frac{d\bold{\tilde{v}}}{dt} = \Lambda_M \textbf{T}^{-1}\bold{\tilde{v}}+ \textbf{T}^{-1}\bold{\tilde{v}}_{\mathrm{in}},
\end{align}
and solve for steady state
\begin{align}
\bold{\tilde{v}}(t) = \int_{-\infty}^t dt' \textbf{T} e^{\Lambda_M (t-t')} \textbf{T}^{-1}\bold{\tilde{v}}_{\mathrm{in}}(t').
\end{align}
One can write the field component (now transforming back from the polaron operators to the bare basis) in a compact form
\begin{align}
a(t)&=\frac{1}{\det \T}\int_{-\infty}^t dt' \left[ \left( \text{T}_{21}\text{T}_{22}e^{\epsilon_{-}(t-t')}-\text{T}_{21}\text{T}_{22}e^{\epsilon_{+}(t-t')}\right)\bold{v}_{\mathrm{in}}^{(1)}(t')+\left(\text{T}_{11}\text{T}_{22}e^{\epsilon_{-}(t-t')}-\text{T}_{12}\text{T}_{21}e^{\epsilon_{+}(t-t')}\right) \bold{v}_{\mathrm{in}}^{(2)}(t') \right]\\\nonumber
&\times \D(t)\Dd(t')
\end{align}
If we want to estimate expectation values, remember that everything is zero except for the driving contribution in the input terms so that (generalizing to many vibrational modes)
\begin{align}
\braket{a(t)}&= \frac {\text{T}_{11}\text{T}_{22}}{\det \T}\sum_{{{\left\{m_\text{k}\right\}}}={0}}^{\infty} \frac{\eta_\text{c} \text{s}_{m_1}^{\lambda_1}...\text{s}_{m_n}^{\lambda_n}}{\gamma_- + \Gamma_{\left\{m_\text{k}\right\}} -i\left(\oml-\omega_- - \nu_{\left\{m_\text{k}\right\}}\right)}-\\ \nonumber
&-\frac {\text{T}_{12}\text{T}_{21}}{\det \T} \sum_{{{\left\{m_k\right\}}}={0}}^{\infty} \frac{\eta_\text{c} \text{s}_{m_1}^{\lambda_1}...\text{s}_{m_n}^{\lambda_n}}{\gamma_+ + \Gamma_{\left\{m_\text{k}\right\}} -i\left(\oml-\omega_+ - \nu_{\left\{m_\text{k}\right\}}\right)}.
\end{align}

\subsection*{Decay of the upper polariton into lower polariton.}
In the single excitation subspace the Jaynes-Cummings part of the Hamiltonian can be diagonalized by defining annihilation operators
for upper and lower polariton (assuming resonance $\omega_c=\omega_e$):
\begin{subequations}
\begin{align}
U&=\frac{1}{\sqrt{2}}\left(a+\sigma\right), \\
L &= \frac{1}{\sqrt{2}}\left(a-\sigma\right).
\end{align}
\end{subequations}
We start with the equations of motion for the bare polaritons
\begin{subequations}
\begin{align}
\frac{dU}{dt} &= -(\gamma_++i\omega_+)U -i\frac{\lambda}{2} \nu  L (b+\bd)+\mathrm{noise},\\
\frac{dL}{dt} &=-(\gamma_-+i\omega_-)L -i\frac{\lambda}{2} \nu  U (b+\bd)+\mathrm{noise},
\end{align}
\end{subequations}
as well as the equation of motion for the population of the lower polariton
\begin{align}
\dot{P}_L=-2\gamma_- P_L+\lambda\nu\Im\braket{L^\dagger U\left(b^\dagger+b\right)},
\end{align}
from which we can see that the term $\Im\braket{L^\dagger U\left(b^\dagger+b\right)}$ is the one responsible for population transfer between the polaritons. Making the notation $\epsilon_+=(\gamma_++i\omega_+)$ and $\epsilon_-=(\gamma_-+i\omega_-)$ we will analyze the decay dynamics of the upper polariton assumming initial conditions $U(0)$ and $L(0)$. The first order solution for upper and lower polaritons are (neglecting the noise terms as they do not contribute to the process):
\begin{subequations}
\begin{align}
U(t)&=U(0) e^{-\epsilon_+ t},\\
L(t)&=L(0)e^{-\epsilon_- t} -i\frac{\lambda}{2}\nu \int_0^t dt' e^{-\epsilon_-\left(t-t'\right)}\left(b(t')+b^\dagger (t')\right)U(0)e^{-\epsilon_+ t'}.
\end{align}
\end{subequations}
with this we can calculate $\braket{L^\dagger U(b^\dagger+b)}$ (using that $\braket{b(t')b^\dagger(t)}=e^{-\left(\Gamma-i\nu\right)(t-t')}$):
\begin{align}
\braket{L^\dagger U(b^\dagger+b)} &= i\frac{\lambda}{2}\nu\int_0^t dt' P_U(0) e^{-(\gamma_--i\omega_-)(t-t')} e^{-(\Gamma-i\nu)(t-t')} e^{-(\gamma_+-i\omega_+)t'} e^{-(\gamma_++i\omega_+)t}\\
&=i\frac{\lambda}{2}\nu\frac{e^{-2\gamma_+t}-e^{-\left(\Gamma+\gamma_-+\gamma_+-i\left(\nu-\omega_+ +\omega_-\right)\right)t}}{\Gamma+\gamma_--\gamma_++i\left(\omega_+-\omega_--\nu\right)}P_U(0).
\end{align}
Assuming $\Gamma\gg\gamma_+,\gamma_-$ and $\gamma_+=\gamma_-$ we can write the population transfer as a rate proportional to the population of the upper polariton:
\begin{align}
\dot{P}_L=-2\gamma_- P_L + \frac{1}{2}\frac{\lambda^2\nu^2\Gamma}{\Gamma^2+\left(\omega_+-\omega_--\nu\right)^2}P_U.
\end{align}

\section{Energy transfer without cavity}
\label{G}
We can write the Langevin equations for dipole-dipole coupled donor/acceptor molecules in the transient regime, assuming some initial population of donor $p_D(0)$ and zero initial population of the acceptor:
\begin{align}
\frac{d\tilde{\s}_D}{dt} &= -\gamma_D \tilde{\s}_D -i \Omega \tilde{\s}_A \D_A \D_D^{\dagger}+ \sqrt{2\gamma_D}\tilde{\s}_D^{\text{in}},\\
\frac{d\tilde{\s}_A}{dt} &= -(\gamma_A-i\Delta) \tilde{\s}_A -i \Omega \tilde{\s}_D \D_D \D_A^\dagger+ \sqrt{2\gamma_A}\tilde{\s}_A^{\text{in}}.
\end{align}
Notice that we have chosen $\Delta=\omega_D-\omega_A>0$. We will perform a perturbative treatment where the evolution of $\tilde{\s}_D(t)$ is in a first approximation given by
\begin{align}
\s_D(t) = \s_D(0)\D_D(t)\D_D^{\dagger}(0)e^{-\gamma_D t}+\sqrt{2\gamma_D}\int_{0}^t dt' e^{-\gamma_D(t-t')}\s_D^{\text{in}}(t')\D_D(t)\D_D^{\dagger}(t')=\s^t_D(t)+\s^s_D(t).
\end{align}
Let us write the hermitian conjugate as well on the two components
\begin{align}
(\s^t_D)^\dagger(t) &= \sd_D(0)\D_D(0)\D_D^{\dagger}(t)e^{-\gamma_D t},\\
(\s^s_D)^\dagger(t) &= \sqrt{2\gamma_D}\int_{0}^t dt' e^{-\gamma_D(t-t')}(\s_D^{\text{in}})^\dagger(t') \D_D(t')\D_D^{\dagger}(t).
\end{align}
The evolution of $\s_A(t)$ is then similarly obtained as:
\begin{align}
\s_A(t) = \s_A(0)\D_A(t)\D_A^\dagger(0)e^{-(\gamma_A-i\Delta) t}+\int_{0}^t dt' e^{-(\gamma_A-i\Delta) (t-t')} \left[-i\Omega \s_D(t')+\sqrt{2\gamma_A}\s_A^{\text{in}}(t')\right]\D_A(t)\D_A^\dagger(t').
\end{align}
We will also split the acceptor operator in a two parts $\s_A(t)=\s^t_A(t)+\s^s_A(t)$. Again the hermitian conjugate on components
\begin{align}
(\s^t_A)^\dagger(t) &= \sd_A(0)\D_A(0)\D_A^{\dagger}(t)e^{-(\gamma_A+i\Delta) t},\\
(\s^s_A)^\dagger(t) &= \int_{0}^t dt' e^{-(\gamma_A+i\Delta) (t-t')} \left[i\Omega \sd_D(t')+\sqrt{2\gamma_A}(\s_A^{\text{in}})^\dagger(t')\right]\D_A(t')\D_A^{\dagger}(t).
\end{align}
From the Eq.~\eqref{LangSimpleAppendix} we see that
\begin{align}
\dot{P}_A = -2 \gamma_A P_A+2\Omega \Im{\braket{\sd_A \s_D}}+\sqrt{2\gamma_A} \braket{\sd_A \s_A^{\text{\text{in}}}+\s_A^{\dagger \text{in}} \s_A}.
\end{align}
The last term containing correlations between acceptor input noise and the evolved acceptor operator does not contribute. The energy transfer is mediated by the $\Im{\braket{\sd_A \s_D}}$ term. We will need to evaluate four terms:
\begin{subequations}
\begin{align}
\braket{(\s^t_A)^\dagger(t) \s^t_D(t)} &= 0,\\
\braket{(\s^t_A)^\dagger(t) \s^s_D(t)} &= 0,\\
\braket{(\s^s_A)^\dagger(t) \s^t_D(t)} &=\braket{\int_{0}^t dt' e^{-(\gamma_A+i\Delta) (t-t')} \left[i\Omega \sd_D(t')+\sqrt{2\gamma_A}(\s_A^{in})^\dagger(t')\right]\D_A(t')\D_A^{\dagger}(t) \s_D(0)\D_D(t)\D_D^{\dagger}(0)e^{-\gamma_D t}}\\\nonumber
&=i\Omega\int_{0}^t dt' e^{-(\gamma_A+i\Delta) (t-t')} e^{-\gamma_D t}\braket{ \left[(\s^t_D)^\dagger(t')+(\s^s_D)^\dagger(t')\right]\s_D(0)\D_D(t)\D_D^{\dagger}(0)}\braket{\D_A(t')\D_A^{\dagger}(t)}\\\nonumber
&=i\Omega\int_{0}^t dt' e^{-(\gamma_A+i\Delta) (t-t')} e^{-\gamma_D t}e^{-\gamma_D t'}\braket{\sd_D(0)\s_D(0)\D_D(0)\D_D^{\dagger}(t')\D_D(t)\D_D^{\dagger}(0)}\braket{\D_A(t')\D_A^{\dagger}(t)}\\
\braket{(\s^s_A)^\dagger(t) \s^s_D(t)} &=0.
\end{align}
\end{subequations}

We can now write
\begin{align}
J=2\Omega \Im \left\{\braket{\sd_A(t)\s_D(t)}\right\}=2\Omega^2 \Re\left\{{\cal{I}}(t)\right\} p_D(0)
\end{align}
and estimate the integral as:
\begin{align}
{\cal{I}}(t)&=\int_{0}^t dt' e^{-(\gamma_A+i\Delta) (t-t')}e^{-\gamma_D(t+t')}e^{-\lambda_D^2e^{-(\Gamma_D+i\nu_D)t'}}e^{\lambda_D^2e^{-(\Gamma_D-i\nu_D)(t-t')}}e^{\lambda_D^2e^{-(\Gamma_D-i\nu_D)t'}}e^{\lambda_A^2e^{-(\Gamma_A-i\nu_A)(t-t')}}e^{-\lambda_D^2}e^{-\lambda_A^2}\\
&=e^{2\lambda_D^2}e^{-(\gamma_A+i\Delta)t} e^{-\gamma_D t} \sum_{n_1,n_2,n_3,n_4}^{\infty}s_{n_1}^{\lambda_D}s_{n_2}^{\lambda_D}s_{n_3}^{\lambda_D}s_{n_4}^{\lambda_A}(-1)^{n_2} e^{-n_1(\Gamma_D-i\nu_D)t}e^{-n_4(\Gamma_A-i\nu_A)t}\times \\\nonumber
& \times\int_{0}^t dt' e^{\left[\gamma_A-\gamma_D+i\Delta+n_1(\Gamma_D-i\nu_D)-n_2(\Gamma_D+i\nu_D)-n_3(\Gamma_D-i\nu_D)+n_4(\Gamma_A-i\nu_A)\right]t'}\\
&= e^{2\lambda_D^2}\sum_{n_1,n_2,n_3,n_4}^{\infty}s_{n_1}^{\lambda_D}s_{n_2}^{\lambda_D}s_{n_3}^{\lambda_D}s_{n_4}^{\lambda_A}(-1)^{n_2}\frac{e^{[-2\gamma_{D} -(n_2+n_3)\Gamma_{D}+i(n_3-n_2)\nu_D]t}-e^{[-(\gamma_A+\gamma_D)-n_1\Gamma_D-n_4\Gamma_A-i(\Delta-n_1\nu_D-n_4\nu_A)]t}}{[\gamma_A-\gamma_D +(n_1-n_2-n_3)\Gamma_D+n_4\Gamma_A + i(\Delta+(n_3-n_2-n_1)\nu_D-n_4\nu_A)]}
\end{align}
Due to the rapid decay of terms harboring $\Gamma_A$, $\Gamma_D$ we can rewrite this term and obtain the expression
\begin{align}
{\cal{I}}(t)&= \sum_{n_1,n_4}^{\infty}s_{n_1}^{\lambda_D}s_{n_4}^{\lambda_A}\frac{e^{-2\gamma_{D}t}-\delta_{0n1}\delta_{0n4}e^{[-(\gamma_A+\gamma_D)-i\Delta]t}}{[\gamma_A-\gamma_D + n_1\Gamma_D+n_4\Gamma_A + i(\Delta-n_1\nu_D-n_4\nu_A)]}.
\end{align}
Ignoring the $n_1 = n_4 =0$ components, the real part of the integral is given by
\begin{align}
\Re\{ {\cal{I}}(t)\}&= \sum_{n_1,n_4}^{\infty}\frac{s_{n_1}^{\lambda_D}s_{n_4}^{\lambda_A}(\gamma_A-\gamma_D + n_1\Gamma_D+n_4\Gamma_A)}{[\gamma_A-\gamma_D + n_1\Gamma_D+n_4\Gamma_A]^2 + [\Delta-n_1\nu_D-n_4\nu_A]^2}e^{-2\gamma_{D}t},
\end{align}
and therefore we obtain
\begin{align}
J &= \sum_{n_1,n_4}^{\infty}\frac{2\Omega^2 s_{n_1}^{\lambda_D}s_{n_4}^{\lambda_A}(\gamma_A-\gamma_D + n_1\Gamma_D+n_4\Gamma_A)}{[\gamma_A-\gamma_D + n_1\Gamma_D+n_4\Gamma_A]^2 + [\Delta-n_1\nu_D-n_4\nu_A]^2}p_De^{-2\gamma_{D}t}.
\end{align}
\section{Energy transfer with cavity}
\label{H}
The Langevin equations for a pair of cavity and dipole-dipole coupled donor/acceptor molecules in the transient regime is given by
\begin{align}
\frac{d\tilde{\s}_D}{dt} &= -\gamma_D \tilde{\s}_D -i \Omega \tilde{\s}_A \D_A \D_D^{\dagger} + g_D a\D_D^{\dagger} + \sqrt{2\gamma_D}\tilde{\s}_D^{\text{in}},\\
\frac{d\tilde{\s}_A}{dt} &= -(\gamma_A-i\Delta) \tilde{\s}_A -i \Omega \tilde{\s}_D \D_D \D_A^{\dagger} + g_A a\D_A^{\dagger} + \sqrt{2\gamma_A}\tilde{\s}_A^{\text{in}},\\
\frac{d a}{dt} &= -(\kappa - i\Delta_c)a -g_D\s_D -g_A\s_A + \sqrt{2\kappa}a^{\text{in}},
\end{align}
where we assume that we have some initial population $P_D(0)$ of the donor and zero initial population in the cavity and of the acceptor. Again we define $\Delta = \omega_D-\omega_A > 0$ and $\Delta_c = \omega_D-\omega_c$.
Following this perturbative ansatz we obtain
\begin{align}
\s_D(t) &= \s_D(0)\D_D(t)\D_D^{\dagger}(0)e^{-\gamma_D t}+\sqrt{2\gamma_D}\int_{0}^t dt' e^{-\gamma_D(t-t')}\s_D^{in}(t')\D_D(t)\D_D^{\dagger}(t')=\s^t_D(t)+\s^s_D(t),\\
\s_A(t) &= \s_A(0)\D_A(t)\D_A^{\dagger}(0)e^{-(\gamma_A-i\Delta) t}+\int_{0}^t dt' e^{-(\gamma_A-i\Delta) (t-t')} \left[-i\Omega \s^t_D(t')+ g_A a(t') +\sqrt{2\gamma_A}\s_A^{in}(t')\right]\D_A(t)\D_A^{\dagger}(t'),\\
a(t) &= a(0)e^{-(\kappa-i\Delta_c)t} + \int_{0}^t dt' e^{-(\kappa-i\Delta_c)(t-t')}\left[-g_D\s^t_D(t') + \sqrt{2\kappa}\s^{in}(t')\right].
\end{align}
Using these result in the differential equation of the acceptor population
\begin{align}
\dot{p}_A = -2 \gamma_A p_A+2\Omega \Im{\braket{\sd_A \s_D}} + 2g_A\Im{\braket{\sd_A a}},
\end{align}
we are left with deriving the terms $\Im{\braket{\sd_A \s_D}}$, $\Im{\braket{\sd_A a}}$.
The dipole-dipole term equates to
\begin{align}
\braket{\sd_A\s_D} &= i\Omega P_D(0)\int^t_0 dt' e^{-(\gamma_A+i\Delta)(t-t')}e^{-\gamma_D(t+t')}\braket{\D_D(0)\D_D^{\dagger}(t')\D_D(t)\D_D^{\dagger}(0)}\braket{\D_A(t')\Dd_{A}(t)}\\
& -g_{A}g_{D}p_D(0)\int^t_0 dt' e^{-(\gamma_A+i\Delta)(t-t')}\int^{t'}_0 dt'' e^{-(\kappa+i\Delta_c)(t'-t'')}e^{-\gamma_D(t+t'')}\braket{\D_D(0)\D_D^{\dagger}(t'')\D_D(t)\Dd_{D}(0)}\nonumber\\
& \times \braket{\D_A(t')\D_A^{\dagger}(t)} \nonumber.
\end{align}
while the cavity term is given by
\begin{align}
\braket{\sd_A a} &= -i\Omega g_D P_D(0) \int^t_0 dt' e^{-(\gamma_A+i\Delta)(t-t')}e^{-\gamma_D t'}\int^{t}_0 dt'' e^{-(\kappa-i\Delta_c)(t-t'')}e^{-\gamma_D t''}\braket{\D_D(0)\D_D^{\dagger}(t')\D_D(t'')\D_D^{\dagger}(0)}\\\nonumber
& \times \braket{\D_A(t')\D_A^{\dagger}(t)}\\
& +g_A g^2_D p_D(0)\int^t_0 dt' e^{-(\gamma_A + i\Delta)(t-t')}\int^{t'}_0 dt'' e^{-(\kappa+i\Delta_c)(t'-t'')}e^{-\gamma_D t''}\int^t_0 dt''' e^{-(\kappa-i\Delta_c)(t-t''')}e^{-\gamma_D t'''}\nonumber\\\nonumber
& \times \braket{\D_D(0)\D_D^{\dagger}(t'')\D_D(t''')\Dd_{D}(0)}\braket{\D_A(t')\D_A^{\dagger}(t)},
\end{align}
where the term with $a^{\dagger}(0)a(0)$ goes to zero. Evaluating the integrals for the dipole-dipole expression results in
\begin{align}
\braket{\sd_A\s_D} &= P_D(0)e^{2\lambda_D^2}\sum_{n_1,n_2,n_3,n_4}^{\infty}s_{n_1}^{\lambda_D}s_{n_2}^{\lambda_D}s_{n_3}^{\lambda_D}s_{n_4}^{\lambda_A}(-1)^{n_2}\\
& \times \left\{ i\Omega\frac{e^{[-2\gamma_{D} -(n_2+n_3)\Gamma_{D}+i(n_3-n_2)\nu_D]t}-e^{[-(\gamma_A+\gamma_D)-n_1\Gamma_D-n_4\Gamma_A-i(\Delta-n_1\nu_D-n_4\nu_A)]t}}{{\scriptstyle [\gamma_A-\gamma_D +(n_1-n_2-n_3)\Gamma_D+n_4\Gamma_A + i(\Delta+(n_3-n_2-n_1)\nu_D-n_4\nu_A)]}} \right. \nonumber\\
& -g_Ag_D\left(\frac{e^{[-2\gamma_D-(n_2+n_3)\Gamma_D-i(n_2-n_3)\nu_D]t}-e^{[-(\gamma_A+\gamma_D)-(n_1\Gamma_D+n_4\Gamma_A)-i(\Delta-n_1\nu_D-n_4\nu_A)]t}}{{\scriptstyle [\gamma_A-\gamma_D +(n_1-n_2-n_3)\Gamma_D+n_4\Gamma_A + i(\Delta+(n_3-n_2-n_1)\nu_D-n_4\nu_A)][\kappa -\gamma_D +(n_1-n_2-n_3)\Gamma_D+i(\Delta_c - (n_1+n_2-n_3)\nu_D)]}} \right. \nonumber\\
& \left.\left. -\frac{e^{[-\kappa - n_1\Gamma_D -i(\Delta_c-n_1\nu_D)]t}-e^{[-(\gamma_A+\gamma_D)]-(n_1\Gamma_D+n_4\Gamma_A)-i(\Delta-n_1\nu_D-n_4\nu_A)]t}}{{\scriptstyle [\gamma_A + n_4\Gamma_A-\kappa+i(\Delta-\Delta_c-n_4\nu_A)][\kappa -\gamma_D +(n_1-n_2-n_3)\Gamma_D+i(\Delta_c - (n_1+n_2-n_3)\nu_D)]}}\right)\right\}\nonumber,
\end{align}
which simplifies to
\begin{align}
\braket{\sd_A\s_D} &= P_D(0)\sum_{n_1,n_4}^{\infty}s_{n_1}^{\lambda_D}s_{n_4}^{\lambda_A} \left\{ i\Omega\frac{e^{-2\gamma_{D}t}-e^{[-(\gamma_A+\gamma_D)-i\Delta]t}\delta_{0n1}\delta_{0n_4}}{ [\gamma_A-\gamma_D +n_1\Gamma_D+n_4\Gamma_A + i(\Delta-n_1\nu_D-n_4\nu_A)]}\right.\\
& -g_Ag_D\left(\frac{e^{-2\gamma_D t}-e^{[-(\gamma_A+\gamma_D)-i\Delta]t}\delta_{0n_1}\delta_{0n_4}}{ [\gamma_A-\gamma_D +n_1\Gamma_D+n_4\Gamma_A + i(\Delta-n_1\nu_D-n_4\nu_A)][\kappa -\gamma_D + n_1\Gamma_D+i(\Delta_c - n_1\nu_D)]} \right. \nonumber\\
& \left.\left. -\frac{\left( e^{[-\kappa-i\Delta_c]t}\delta_{0n_1}-e^{[-(\gamma_A+\gamma_D)-i\Delta]t}\delta_{0n_1}\delta_{0n_4}\right)}{ [\gamma_A + n_4\Gamma_A-\kappa+i(\Delta-\Delta_c-n_4\nu_A)][\kappa -\gamma_D +n_1\Gamma_D+i(\Delta_c - n_1\nu_D)]}\right)\right\}\nonumber,
\end{align}
if we take into consideration that terms that contain $\Gamma_D$ or $\Gamma_A$ decay rapidly.
By focussing on the terms that decay with $e^{-2\gamma_D t}$ and ignoring all oscillatory terms we finally obtain
\begin{align}
J = 2\Omega\Im{\braket{\sd_A\s_D}} &\approx P_D(0)e^{-2\gamma_D t}\sum_{n_1,n_4}^{\infty}s_{n_1}^{\lambda_D}s_{n_4}^{\lambda_A} \left\{\left(\frac{2\Omega^2 [\gamma_A-\gamma_D+n_1\Gamma_D+n_4\Gamma_A]}{ [(\gamma_A-\gamma_D +n_1\Gamma_D+n_4\Gamma_A)^2 + (\Delta-n_1\nu_D-n_4\nu_A)^2]}\right)\right.\\
& - \left.\left(\frac{2\Omega g_Ag_D [(\Delta-n_1\nu_D-n_4\nu_A)(\kappa -\gamma_D + n_1\Gamma_D)+(\Delta_c - n_1\nu_D)(\gamma_A-\gamma_D +n_1\Gamma_D+n_4\Gamma_A)]}{ [(\gamma_A-\gamma_D +n_1\Gamma_D+n_4\Gamma_A)^2 + (\Delta-n_1\nu_D-n_4\nu_A)^2][(\kappa -\gamma_D + n_1\Gamma_D)^2+(\Delta_c - n_1\nu_D)^2]}\right)\right\}\nonumber
\end{align}
Evaluating the integrals for the cavity induced term results in
\begin{align}
\braket{\sd_A a} &= P_D(0)e^{4\lambda_D^2}\sum_{n_1,\dots,n_6}^{\infty}s_{n_1}^{\lambda_D}s_{n_2}^{\lambda_D}s_{n_3}^{\lambda_D}s_{n_5}^{\lambda_D}s_{n_6}^{\lambda_D}s_{n_4}^{\lambda_A}(-1)^{n_2+n_5}\\\nonumber
\times & \left\{-i\Omega g_D\left( \frac{e^{[-2\gamma_D-(n_2+n_3+n_5+n_6)\Gamma_D+i(n_2-n_3-n_5+n_6)\nu_D]t}-e^{[-\gamma_A-\kappa-n_4\Gamma_A-i(\Delta-\Delta_c-n_4\nu_A)]t}}{{\scriptstyle [\gamma_A+\kappa-2\gamma_D-(n_2+n_3+n_5+n_6)\Gamma_D+n_4\Gamma_A+i(\Delta-\Delta_c+(n_2-n_3-n_5+n_6)\nu_D-n_4\nu_A)][\kappa-\gamma_D+(n_1-n_2-n_3)\Gamma_D-i(\Delta_c+(n_3-n_2-n_1)\nu_D)]}} \right.\right. \\\nonumber
- & \frac{\left( e^{[-\kappa-\gamma_D-(n_1+n_5+n_6)\Gamma_D+i(\Delta_c+(n_6-n_5-n_1)\nu_D)]t}-e^{[-\gamma_A-\kappa-n_4\Gamma_A-i(\Delta-\Delta_c-n_4\nu_A)]t}\right)}{{\scriptstyle [\gamma_A-\gamma_D-(n_1+n_5+n_6)\Gamma_D+n_4\Gamma_A+i(\Delta+(n_6-n_5-n_1)\nu_D-n_4\nu_A)][\kappa-\gamma_D+(n_1-n_2-n_3)\Gamma_D-i(\Delta_c+(n_3-n_2-n_1)\nu_D)]}}\\\nonumber
+ &  \frac{e^{[-2\gamma_D-(n_2+n_3+n_5+n_6)\Gamma_D-i(n_2-n_3-n_5+n_6)\nu_D]t}-e^{[-\gamma_A-\kappa-n_4\Gamma_A-i(\Delta-\Delta_c-n_4\nu_A)]t}}{{\scriptstyle [\gamma_A+\kappa-2\gamma_D-(n_2+n_3+n_5+n_6)\Gamma_D+n_4\Gamma_A+i(\Delta-\Delta_c-(n_2-n_3-n_5+n_6)\nu_D-n_4\nu_A)][\gamma_A-\gamma_D+(n_1-n_2-n_3)\Gamma_D+n_4\Gamma_A+i(\Delta+(n_3-n_2-n_1)\nu_D-n_4\nu_A)]}}\\\nonumber
-& \left. \frac{\left(e^{[-(\gamma_A+\gamma_D)-(n_1+n_5+n_6)\Gamma_D-n_4\Gamma_A-i(\Delta-(n_1+n_5-n_6)\nu_D-n_4\nu_A)]t}-e^{[-\gamma_A-\kappa-n_4\Gamma_A-i(\Delta-\Delta_c-n_4\nu_A)]t}\right)}{{\scriptstyle [\kappa-\gamma_D-(n_1+n_5+n_6)\Gamma_D-i(\Delta_c-(n_1+n_5-n_6)\nu_D)][\gamma_A-\gamma_D+(n_1-n_2-n_3)\Gamma_D+n_4\Gamma_A+i(\Delta+(n_3-n_2-n_1)\nu_D-n_4\nu_A)]}}\right)\\\nonumber
+&g_A g^2_D\left(\frac{e^{[-2\gamma_D-(n_2+n_3+n_5+n_6)\Gamma_D+i(n_2-n_3-n_5+n_6)\nu_D]t}-e^{[-\gamma_A-\kappa-n_4\Gamma_A-i(\Delta-\Delta_c-n_4\nu_A)]t}}{{\scriptstyle [\kappa+\gamma_A-2\gamma_D-(n_2+n_3+n_5+n_6)\Gamma_D+n_4\Gamma_A+i(\Delta-\Delta_c+(n_2-n_3-n_5+n_6)\nu_D-n_4\nu_A)][\dots (1)\dots ][\dots (2) \dots ]}} \right.\\\nonumber
- & \frac{\left(e^{[-\kappa-\gamma_D+(n_1-n_2-n_3)\Gamma_D-i(\Delta_c+(n_1+n_2-n_3)\nu_D)]t}-e^{[-\gamma_A-\kappa-n_4\Gamma_A-i(\Delta-\Delta_c-n_4\nu_A)]t}\right)}{{\scriptstyle [\gamma_A-\gamma_D+(n_1-n_2-n_3)\Gamma_D+n_4\Gamma_A + i(\Delta-2\Delta_c+(n_1+n_2-n_3)\nu_D-n_4\nu_A)][\dots (1)\dots][\dots (2)\dots]}} \\\nonumber
- & \frac{\left( e^{[-\kappa-\gamma_D-(n_1+n_5+n_6)\Gamma_D+i(\Delta_c+(n_6-n_5-n_1)\nu_D)]t}-e^{[-\gamma_A-\kappa-n_4\Gamma_A-i(\Delta-\Delta_c-n_4\nu_A)]t}\right)}{{\scriptstyle [\gamma_A-\gamma_D+(n_1+n_5+n_6)\Gamma_D+n_4\Gamma_A+i(\Delta+(n_6-n_5-n_1)\nu_D-n_4\nu_A)][\dots (1)\dots][\dots (2)\dots]}}\\\nonumber
+ & \frac{e^{-2\kappa t}-e^{[-\gamma_A-\kappa-n_4\Gamma_A-i(\Delta-\Delta_c-n_4\nu_A)]t}}{{\scriptstyle [\gamma_A-\kappa + n_4\Gamma_A+i(\Delta-\Delta_c-n_4\nu_A)][\dots (1)\dots][\dots (2)\dots]}}\\\nonumber
+ & \frac{e^{[-2\gamma_D-(n_2+n_3+n_5+n_6)\Gamma_D-i(n_2-n_3-n_5+n_6)\nu_D]t}-e^{[-\gamma_A-\kappa-n_4\Gamma_A-i(\Delta-\Delta_c-n_4\nu_A)]t}}{{\scriptstyle [\gamma_A-\gamma_D+(n_1-n_2-n_3)\Gamma_D+n_4\Gamma_A+i(\Delta+(n_3-n_2-n_1)\nu_D-n_4\nu_A)][\dots (3)\dots ][\dots (4) \dots]}}\\\nonumber
- & \frac{\left(e^{[-\kappa-\gamma_D-(n_1+n_5+n_6)\Gamma_D-i(\Delta_c-(n_1+n_5-n_6)\nu_D)]t}-e^{[-\gamma_A-\kappa-n_4\Gamma_A-i(\Delta-\Delta_c-n_4\nu_A)]t}\right)}{{\scriptstyle [\gamma_A-\kappa+n_4\Gamma_A+i(\Delta-\Delta_c-n_4\nu_A)][\dots (3)\dots ][\dots (4) \dots]}}\\\nonumber
- & \frac{\left(e^{[-\kappa-\gamma_D+(n_1-n_2-n_3)\Gamma_D+i(\Delta_c+(n_3-n_2-n_1)\nu_D)]t}-e^{[-\gamma_A-\kappa-n_4\Gamma_A-i(\Delta-\Delta_c-n_4\nu_A)]t}\right)}{{\scriptstyle [\gamma_A-\gamma_D+(n_1-n_2-n_3)\Gamma_D+n_4\Gamma_A+i(\Delta+(n_3-n_2-n_1)\nu_D-n_4\nu_A)][\dots (3)\dots ][\dots (4) \dots]}}\\\nonumber
+ & \left.\left. \frac{\left(e^{-2\kappa t}-e^{[-\gamma_A-\kappa-n_4\Gamma_A-i(\Delta-\Delta_c-n_4\nu_A)]t} \right)}{{\scriptstyle [\gamma_A-\kappa+n_4\Gamma_A+i(\Delta-\Delta_c-n_4\nu_A)][\dots (3)\dots ][\dots (4) \dots]}} \right) \right\},
\end{align}
with
\begin{align}
[\dots (1)\dots ] &= [\kappa -\gamma_D+(n_1-n_2-n_3)\Gamma_D-i(\Delta_c+(n_3-n_2-n_1)\nu_D)] = [\dots (3)\dots ]^{*}\\
[\dots (2)\dots ] &= [\kappa -\gamma_D+(n_1+n_5+n_6)\Gamma_D+i(\Delta_c-(n_1+n_5-n_6)\nu_D)] = [\dots (4) \dots]^{*}.
\end{align}
Again by using the fact that $\Gamma_D$ or $\Gamma_A$ decay rapidly, we obtain
\begin{align}
\braket{\sd_A a} &= P_D(0)\sum_{n_1,n_4}^{\infty}s_{n_1}^{\lambda_D}s_{n_4}^{\lambda_A}\left\{-i\Omega g_D \left(\frac{e^{-2\gamma_D t}-e^{[-\gamma_A-\kappa-i(\Delta-\Delta_c)]t}\delta_{0n_4}}{{\scriptstyle [\gamma_A+\kappa-2\gamma_D+n_4\Gamma_A+i(\Delta-\Delta_c-n_4\nu_A)][\kappa-\gamma_D+n_1\Gamma_D-i(\Delta_c-n_1\nu_D)] }} \right. \right. \\\nonumber
- & \frac{\left(e^{[-\kappa-\gamma_D+i\Delta_c]t}\delta_{0n_1}-e^{[-\gamma_A-\kappa-i(\Delta-\Delta_c)]t}\delta_{0n_4} \right)}{{\scriptstyle [\gamma_A-\gamma_D-n_1\Gamma_D+n_4\Gamma_A+i(\Delta-n_1\nu_D-n_4\nu_A)][\kappa-\gamma_D+n_1\Gamma_D-i(\Delta_c-n_1\nu_D)]}} \\\nonumber
+ & \frac{e^{-2\gamma_D t}-e^{[-\gamma_A-\kappa-i(\Delta-\Delta_c)]t}\delta_{0n_4}}{{\scriptstyle [\gamma_A+\kappa-2\gamma_D+n_4\Gamma_A+i(\Delta-\Delta_c-n_4\nu_A)][\gamma_A-\gamma_D+n_1\Gamma_D+n_4\Gamma_A+i(\Delta-n_1\nu_D-n_4\nu_A)] }} \\\nonumber
- & \left. \frac{\left(e^{[-(\gamma_A+\gamma_D)-i\Delta]t}\delta_{0n_1}\delta_{0n_4}-e^{[-\gamma_A-\kappa-i(\Delta-\Delta_c)]t}\delta_{0n_4} \right)}{{\scriptstyle [\gamma_A-\gamma_D+n_1\Gamma_D+n_4\Gamma_A+i(\Delta-n_1\nu_D-n_4\nu_A)][\kappa-\gamma_D-n_1\Gamma_D-i(\Delta_c-n_1\nu_D)]}} \right) \\\nonumber
+ & g_A g^2_D\left(\frac{e^{-2\gamma_D t}-e^{[-\gamma_A-\kappa-i(\Delta-\Delta_c)]t}\delta_{0n_4}}{{\scriptstyle [\kappa+\gamma_A-2\gamma_D+n_4\Gamma_A+i(\Delta-\Delta_c-n_4\nu_A)][\kappa-\gamma_D+n_1\Gamma_D-i(\Delta_c-n_1\nu_D)][\kappa-\gamma_D-n_1\Gamma_D+i(\Delta_c-n_1\nu_D)]}} \right.\\\nonumber
- & \frac{\left(e^{[-\kappa-\gamma_D-i\Delta_c]t}\delta_{0n_1}-e^{[-\gamma_A-\kappa-i(\Delta-\Delta_c)]t}\delta_{0n_4} \right)}{{\scriptstyle [\gamma_A-\gamma_D+n_1\Gamma_D+n_4\Gamma_A+i(\Delta-2\Delta_c+n_1\nu_D-n_4\nu_A)][\kappa-\gamma_D+n_1\Gamma_D-i(\Delta_c-n_1\nu_D)][\kappa-\gamma_D-n_1\Gamma_D+i(\Delta_c-n_1\nu_D)]}} \\\nonumber
- & \frac{\left(e^{[-\kappa-\gamma_D+i\Delta_c]t}\delta_{0n_1}-e^{[-\gamma_A-\kappa-i(\Delta-\Delta_c)]t}\delta_{0n_4} \right)}{{\scriptstyle [\gamma_A-\gamma_D+n_1\Gamma_D+n_4\Gamma_A+i(\Delta-n_1\nu_D-n_4\nu_A)][\kappa-\gamma_D+n_1\Gamma_D-i(\Delta_c-n_1\nu_D)][\kappa-\gamma_D-n_1\Gamma_D+i(\Delta_c-n_1\nu_D)]}} \\\nonumber
+ & \frac{e^{-2\kappa t}-e^{[-\gamma_A-\kappa-i(\Delta-\Delta_c)]t}\delta_{0n_4}}{{\scriptstyle [\gamma_A-\kappa+n_4\Gamma_A+i(\Delta-\Delta_c-n_4\nu_A)][\kappa-\gamma_D+n_1\Gamma_D-i(\Delta_c-n_1\nu_D)][\kappa-\gamma_D-n_1\Gamma_D+i(\Delta_c-n_1\nu_D)]}}\\\nonumber
+ & \frac{e^{-2\gamma_D t}-e^{[-\gamma_A-\kappa-i(\Delta-\Delta_c)]t}\delta_{0n_4}}{{\scriptstyle [\gamma_A-\gamma_D+n_1\Gamma_D+n_4\Gamma_A+i(\Delta-n_1\nu_D-n_4\nu_A)][\kappa-\gamma_D+n_1\Gamma_D+i(\Delta_c-n_1\nu_D)][\kappa-\gamma_D-n_1\Gamma_D-i(\Delta_c-n_1\nu_D)]}}\\\nonumber
- & \frac{\left(e^{[-\kappa-\gamma_D-i\Delta_c]t}\delta_{0n_1}-e^{[-\gamma_A-\kappa-i(\Delta-\Delta_c)]t}\delta_{0n_4} \right)}{{\scriptstyle [\gamma_A-\kappa+n_4\Gamma_A+i(\Delta-\Delta_c-n_4\nu_A)][\kappa-\gamma_D+n_1\Gamma_D+i(\Delta_c-n_1\nu_D)][\kappa-\gamma_D-n_1\Gamma_D-i(\Delta_c-n_1\nu_D)]}} \\\nonumber
- & \frac{\left(e^{[-\kappa-\gamma_D+i\Delta_c]t}\delta_{0n_1}-e^{[-\gamma_A-\kappa-i(\Delta-\Delta_c)]t}\delta_{0n_4} \right)}{{\scriptstyle [\gamma_A-\gamma_D+n_1\Gamma_D+n_4\Gamma_A+i(\Delta-n_1\nu_D-n_4\nu_A)][\kappa-\gamma_D+n_1\Gamma_D+i(\Delta_c-n_1\nu_D)][\kappa-\gamma_D-n_1\Gamma_D-i(\Delta_c-n_1\nu_D)]}}\\\nonumber
+ & \left. \left. \frac{e^{-2\kappa t}-e^{[-\gamma_A-\kappa-i(\Delta-\Delta_c)]t}\delta_{0n_4}}{{\scriptstyle [\gamma_A-\kappa+n_{4}\Gamma_A+i(\Delta-\Delta_c-n_4\nu_A)][\kappa-\gamma_D+n_1\Gamma_D+i(\Delta_c-n_1\nu_D)][\kappa-\gamma_D-n_1\Gamma_D-i(\Delta_c-n_1\nu_D)]}} \right) \right\}.
\end{align}
By focussing on the terms that decay with $e^{-2\gamma_D t}$ or $e^{-2\kappa t}$ and ignoring all oscillatory terms we obtain
\begin{align}
J_A  & = 2g_A\Im{\braket{\sd_A a}} \approx \sum_{n_1,n_4}^{\infty}s^{\lambda_D}_{n_1}s^{\lambda_A}_{n_4}\left[\left\{-2\Omega g_A g_D\left(\frac{{\scriptstyle [(\gamma_A+\kappa-2\gamma_D+n_4\Gamma_A)(\kappa-\gamma_D+n_1\Gamma_D)+(\Delta-\Delta_c-n_4\nu_A)(\Delta_c-n_1\nu_D)]}}{{\scriptstyle [(\gamma_A+\kappa-2\gamma_D+n_4\Gamma_A)^2+(\Delta-\Delta_c-n_4\nu_A)^2][(\kappa-\gamma_D+n_1\Gamma_D)^2+(\Delta_c-n_1\nu_D)^2]}} \right.\right.\right.\\\nonumber
& \left. + \frac{{\scriptstyle [(\gamma_A+\kappa-2\gamma_D+n_4\Gamma_A)(\gamma_A-\gamma_D+n_1\Gamma_D+n_4\Gamma_A) - (\Delta-\Delta_c-n_4\nu_A)(\Delta-n_1\nu_D-n_4\nu_A)]}}{{\scriptstyle [(\gamma_A+\kappa-2\gamma_D+n_4\Gamma_A)^2+(\Delta-\Delta_c-n_4\nu_A)^2][(\gamma_A-\gamma_D+n_1\Gamma_D+n_4\Gamma_A)^2+(\Delta-n_1\nu_D-n_4\nu_A)^2] }}\right)\\\nonumber
& + 2g_A^2 g_D^2\left(\frac{{\scriptstyle \{(\Delta-\Delta_c-n_4\nu_A)[((\kappa-\gamma_D)^2-(n_1\Gamma_D)^2)+(\Delta_c-n_1\nu_D)^2] + (\kappa+\gamma_A-2\gamma_D+n_4\Gamma_A)[2(\Delta_c-n_1\nu_D)n_1\Gamma_D] \}}}{{\scriptstyle [(\kappa+\gamma_A-2\gamma_D+n_4\Gamma_A)^2+(\Delta-\Delta_c-n_4\nu_A)^2][(\kappa-\gamma_D+n_1\Gamma_D)^2+(\Delta_c-n_1\nu_D)^2][(\kappa-\gamma_D-n_1\Gamma_D)^2+(\Delta_c-n_1\nu_D)^2]}} \right. \\\nonumber
& \left. \left. + \frac{{\scriptstyle \{(\Delta-n_1\nu_D-n_4\nu_A)[((\kappa-\gamma_D)^2-(n_1\Gamma_D)^2)+(\Delta_c-n_1\nu_D)^2] - (\gamma_A-\gamma_D+n_1\Gamma_D+n_4\Gamma_A)[2(\Delta_c-n_1\nu_D)n_1\Gamma_D] \}}}{{\scriptstyle [(\gamma_A-\gamma_D+n_1\Gamma_D+n_4\Gamma_A)^2+(\Delta-n_1\nu_D-n_4\nu_A)^2][(\kappa-\gamma_D+n_1\Gamma_D)^2+(\Delta_c-n_1\nu_D)^2][(\kappa-\gamma_D-n_1\Gamma_D)^2+(\Delta_c-n_1\nu_D)^2]}} \right)\right\}p_D(0)e^{-2\gamma_D t}\\\nonumber
& + \left\{2g_A^2 g_D^2\left( \frac{{\scriptstyle \{(\Delta-\Delta_c-n_4\nu_A)[((\kappa-\gamma_D)^2-(n_1\Gamma_D)^2)+(\Delta_c-n_1\nu_D)^2] + (\gamma_A-\kappa+n_4\Gamma_A)[2(\Delta_c-n_1\nu_D)n_1\Gamma_D] \}}}{{\scriptstyle [(\gamma_A-\kappa+n_4\Gamma_A)^2+(\Delta-\Delta_c-n_4\nu_A)^2][(\kappa-\gamma_D+n_1\Gamma_D)^2+(\Delta_c-n_1\nu_D)^2][(\kappa-\gamma_D-n_1\Gamma_D)^2+(\Delta_c-n_1\nu_D)^2]}} \right.\right. \\\nonumber
& \left.\left. \left. + \frac{{\scriptstyle \{(\Delta-\Delta_c-n_4\nu_A)[((\kappa-\gamma_D)^2-(n_1\Gamma_D)^2)+(\Delta_c-n_1\nu_D)^2] - (\gamma_A-\kappa+n_{4}\Gamma_A)[2(\Delta_c-n_1\nu_D)n_1\Gamma_D] \}}}{{\scriptstyle [(\gamma_A-\kappa+n_{4}\Gamma_A)^2+(\Delta-\Delta_c-n_4\nu_A)^2][(\kappa-\gamma_D+n_1\Gamma_D)^2+(\Delta_c-n_1\nu_D)^2][(\kappa-\gamma_D-n_1\Gamma_D)^2+(\Delta_c-n_1\nu_D)^2]}} \right) \right\} P_D(0)e^{-2\kappa t}\right]
\end{align}

\end{document}